\documentclass[prd,a4paper,aps,showpacs,
               nofootinbib,nobibnotes,superscriptaddress]
              {revtex4}
\usepackage{epsfig}
\usepackage{amssymb}

\newcommand{\bn}{{\bf n}} 
 
\newcommand{\A}{{\cal A}} 
\newcommand{\BB}{{\cal B}} 
\newcommand{\HH}{{\cal H}} 
\newcommand{\cd}{\cdot} 
 
\newcommand{\al}{\alpha} 
 
\newcommand{\de}{\delta} 
\newcommand{\De}{\Delta} 
\newcommand{\ep}{\epsilon} 
\newcommand{\ga}{\gamma} 
\newcommand{\Ga}{\Gamma}

\newcommand{\la}{\lambda} 
\newcommand{\Om}{\Omega}

\newcommand{\Th}{\Theta}

\newcommand{\ra}{\rightarrow}

\newcommand{\be}{\begin{equation}} 
\newcommand{\ee}{\end{equation}} 
\newcommand{\gsim}{\gtrsim} 
\newcommand{\lsim}{\lesssim} 
\newcommand{\bea}{\begin{eqnarray}} 
\newcommand{\eea}{\end{eqnarray}} 
\newcommand{\bean}{\begin{eqnarray*}} 
\newcommand{\eean}{\end{eqnarray*}}

\newcommand{\bk}{{\mathbf k}} 
\newcommand{\bp}{{\mathbf p}} 
\newcommand{\bq}{{\mathbf q}} 
\newcommand{\bB}{{\mathbf B}} 
\newcommand{\by}{{\mathbf y}} 
\newcommand{\bx}{{\mathbf x}} 
\newcommand{\bv}{{\mathbf v}} 
\newcommand{\boe}{{\mathbf e}} 
\newcommand{\ie}{{\em i.e. }} 
\newcommand{\eg}{{\em e.g. }}

\begin{document} 
%\draft 
\title{The Cosmic Microwave Background and Helical Magnetic Fields:\\ 
  the tensor mode} 
\author{Chiara Caprini} 
\email{caprini@astro.ox.ac.uk} 
\affiliation{Department of Astrophysics, Denys Wilkinson Building, 
  Keble road, Oxford OX1 3RH, UK} 
 
\author{Ruth Durrer} 
\email{ruth.durrer@physics.unige.ch} 
\affiliation{D\'epartement de Physique Th\'eorique, Universit\'e de 
  Gen\`eve, 24 quai Ernest Ansermet, CH--1211 Gen\`eve 4, Switzerland} 
 
\author{Tina Kahniashvili} 
\email{tinatin@amorgos.unige.ch} 
\affiliation{Department of Physics and Astronomy, Rutgers NJ State
  University, 136, Frelinghuysen RD., Piscataway, NJ, 08854-8019, USA
\\  and\\
Center for Plasma Astrophysics, Abastumani Astrophysical Observatory,
2A, Kazbegi ave., Tbilisi, 380060, Georgia} 

\date{\today} 
 
\begin{abstract} 
 
We study the effect of a possible helicity component of a primordial
magnetic field on the tensor part of the cosmic microwave background temperature
anisotropies and polarization. We give analytical 
approximations for the tensor contributions induced by helicity,
discussing their amplitude and spectral index in dependence of the power
spectrum of the primordial magnetic field. We find that an helical
magnetic field creates a parity odd component of gravity waves inducing
parity odd polarization signals. However, only if the
magnetic field is close to scale invariant and if its helical
part is close to maximal, the effect is sufficiently large to be
observable. We also discuss the implications of causality on the 
magnetic field spectrum. 
\end{abstract}

 
\maketitle

\section{Introduction} 
\label{sec:intro} 
The observed Universe is permeated with large scale coherent magnetic 
fields. It is still under debate whether these magnetic fields have 
been created by charge separation processes in the late Universe, or 
whether primordial seed fields are needed. Recently, it 
has been proposed~\cite{vachaspati01} that also `helical'
magnetic fields,  \ie fields with a
non-vanishing component in the direction of the current, ${\bf B\cd 
(\nabla\times B)} \neq 0$, could be produced \eg during the 
electroweak phase transition (see also \cite{brandenburg02}). 
 
Extended studies have already investigated effects of stochastic 
magnetic fields with vanishing helicity on the cosmic microwave
background (CMB) (see~\cite{jedamzik98,durrer00,grasso01,mack02} and
others). In a seminal paper~\cite{pogosian02}, Pogosian and
collaborators have  
investigated the possibility that a helical magnetic field can induce 
correlations between the temperature anisotropy and the $B$ mode CMB 
polarization. 
 
In this paper we want to go beyond that work. We determine all the 
effects on the CMB induced by a helical magnetic field. We shall 
actually show that, contrary to the statement in 
Ref.~\cite{pogosian02}, a helical component also introduces pure CMB 
anisotropies and polarization. But of course its most remarkable 
effect is the above mentioned correlation of temperature anisotropy 
and $B$ polarization. 
We shall show that also a correlation between $E$ and $B$ polarization 
is induced. 
 
In this paper we discuss only the tensor mode, gravitational waves, 
since the calculations for this case are  simplest. Even if the
resulting observational effects are small and may not be 
detectable, we find it 
interesting since it is completely new and contains several surprising 
elements. Furthermore, a fluid vorticity field or non parity invariant initial
spectrum of gravitational waves produced during inflation could induce very 
similar effects; in that sense our results are more 
generic than their derivation. 
 
In the next section, we discuss the magnetic field spectrum and define 
its symmetric and helical contributions. Then we compute the tensor 
component of the magnetic field energy momentum tensor which acts as a 
source for gravity waves. In Section~IV we determine the induced 
gravity wave spectrum which also has a symmetric and a helical 
contribution. In Section~V we compute the induced CMB temperature 
anisotropy and polarization spectra as well as the above mentioned 
correlations. Finally, we discuss our results and draw some conclusions. 
The paper is complemented by an appendix where details of 
calculations and tests of some approximations can be found. 
 
\section{The magnetic field spectrum} 
\label{sec:magnetic field} 
 
We consider a primordial stochastic magnetic field created before 
equality, during the radiation-dominated epoch (or earlier). 
During this period of the evolution of the Universe, 
the conductivity of the primordial plasma on scales larger 
than the Silk scale $\lambda > \lambda_S$ is very high, effectively 
infinite \cite{ahonen96}. Hence, the `frozen-in' 
condition holds, $\mathbf E=-\mathbf{v}\times \mathbf{B}$, 
where ${\mathbf v}$ is the 
plasma flux velocity, ${\mathbf E}$ is the electric field induced 
by plasma motions and ${\mathbf B}$ is the magnetic field.  Moreover, 
large scale magnetic fields always induce anisotropic stresses, so 
that their energy density $B^2/8\pi$ must be a small perturbation, 
in order not to break the isotropy of the Friedmann Robertson 
Walker background.  This allows us to apply linear 
perturbation theory. Both, the magnetic field energy and the plasma peculiar 
velocity are treated as first order perturbations; 
consequently, the energy density of the induced electric field 
will be $3$rd order  in perturbations theory, and can be neglected. 
Also terms $E_iB_j$ are of second order and therefore neglected.
 
At sufficiently large scales, it is possible to neglect the effects of 
back reaction of the fluid on the evolution of the magnetic field: 
the time dependence decouples from the 
spatial structure, and, due to flux conservation, the magnetic 
field evolves like ${\mathbf B}(\eta,{\mathbf 
  x})={\mathbf B}(\eta_0, {\mathbf x})/a(\eta)^2$, where we use the
normalization $a(\eta_0)=1$ and a subscript $0$ denotes today. 
At smaller scales however, the interaction 
between the fluid and the magnetic field becomes important, leading 
mainly to two effects: on intermediated scale, the plasma undergoes 
Alfv\'en oscillations, and $B^2 (k) \rightarrow B^2 (k)\cos^2(v_A k \eta)$ 
(where $v_A^2=B^2/(4\pi(\rho+p))$ 
is the Alfv\'en velocity, here $B$ is the field averaged over a scale of 
order $v_A\eta$); on very small scales, the field is exponentially 
damped due to shear viscosity 
\cite{jedamzik98,subramanian98b,subramanian98a,durrer00}. 
As in Ref.~\cite{durrer00}, we will account for this damping by 
introducing an ultraviolet cutoff at wavenumber $k_D(\eta)$ in the spectrum of 
$\mathbf{B}$ (see also \cite{mack02}). 
 
Following Refs. \cite{vachaspati01,pogosian02}, we introduce an helicity 
component $A(k)$ in   the magnetic field two point correlation function: 
\begin{equation} 
\langle B_j({\mathbf k})B^{*}_l({\mathbf k'})\rangle 
=\frac{(2\pi)^3}{2} \delta({\mathbf k}-{\mathbf k'}) [P_{jl} S(k)   + 
i \epsilon_{jlm} \hat{k}_m A(k)]~, 
\label{spectrum} 
\end{equation} 
where $S(k)$ and $A(k)$ are respectively the symmetric and helical part 
of the magnetic field power spectrum. 
$P_{ij}\equiv\delta_{ij}-\hat{k}_i\hat{k}_j$ 
is the usual transverse plane projector satisfying the conditions 
$P_{ij}P_{jk}=P_{ik}$, $P_{ij}\hat{k}_j=0$, $\epsilon_{ijl}$ is the totally 
antisymmetric tensor, and $\hat{k}_i=k_i/k$. We use the Fourier 
transformation convention 
\begin{equation} 
B_j({\mathbf k}) = \int d^3x\, 
\exp(i{\mathbf k}\cdot {\mathbf x}) B_j({\mathbf x}), 
~~~~~~~~~~~~~~~~~~ 
B_j({\mathbf x}) = {1 \over {(2\pi)^3}} \int d^3k\, 
\exp(-i{\mathbf k}\cdot {\mathbf x}) B_j({\mathbf k})~. 
\end{equation} 
For simplicity, as in Refs.~\cite{durrer00,mack02} and others, we 
shall assume that the magnetic field is a Gaussian random field. Then 
all the statistical information is contained in the two-point 
correlation function and the higher moments can be obtained via Wick's 
theorem. 
 
As explained in Ref.~\cite{pogosian02}, the
magnetic field helicity is determined by
$\langle {\mathbf B} \cdot (\nabla \times {\mathbf B}) \rangle $. 
For a better physical understanding of the effects which this new
helicity term  has on CMB anisotropies, it is useful to introduce
the orthonormal `helicity basis'  
$({\mathbf e}^{+},~{\mathbf e}^{-},~{\mathbf e}_3={\hat\bk})$ 
 (see also \cite{hu96,pogosian02}), where 
\be 
{\mathbf e}^{\pm }({\bf k}) 
=-\frac{i}{\sqrt{2}}({\mathbf e}_1 \pm  i{\mathbf e}_2)~, 
\label{vector-basis} 
\ee 
and $({\mathbf e}_1,~{\mathbf e}_2,~{\mathbf e}_3={\hat \bk})$ 
form a right-handed orthonormal basis with ${\mathbf e}_2 ={ 
  \hat \bk}\times {\mathbf e}_1$. Under the transformation ${\mathbf k} 
 \ra -{\bk}$ we choose $\mathbf e_2$ to change sign while 
$\mathbf e_1$ remains invariant. 
The basis $(\boe^+,\boe^-,\hat\bk)$ has the following properties: 
${\mathbf e}^{\pm}\cdot {\mathbf e}^{\mp}=-1$,~ 
${\mathbf e}^{\pm}\cdot {\mathbf e}^{\pm}=0$,~ 
and 
${\mathbf e}^{\pm }({\bk})={\mathbf e}^{\mp}(-{\bk})$, as 
well as 
$i\hat{\bk} \times {\mathbf e}^{\pm}=\pm {\mathbf e}^{\pm}$.~ 
The components of a vector with 
respect to this basis will be indicated by a superscript $\pm$. For a 
fixed ($k$-independent) basis we will 
instead use the usual Latin letters as indices. An arbitrary
transverse vector ${\bf v}$ can be decomposed as $\bv = v^+\boe^+ + v^-\boe^-$.
Here $v^+$ is the positive helicity component and $v^-$ is the negative 
 helicity component.

With the definition (\ref{spectrum}), and the reality condition 
$(B^{\pm }({\bf k}))^* = -B^{\pm}(-{\mathbf k})$, we obtain the 
connection between the power spectra $S(k)$, $A(k)$ and the magnetic field 
components in the new basis: 
\bea 
-\langle B^{+}{(\mathbf k)} B^{+}(-{\mathbf k^\prime}) + 
 B^{-}({\mathbf k}) B^{-}(-{\mathbf k^\prime}) \rangle &=&(2\pi)^3 S(k) 
\delta ({\mathbf k} - {\mathbf k^\prime})~,\label{B-vector-basis}\\ 
\langle B^{+}({\mathbf k}) B^{+}(-{\mathbf k^\prime)} - 
 B^{-}({\mathbf k}) B^{-}(-{\mathbf k^\prime})\rangle &=& (2\pi)^3 A(k) 
\delta ({\mathbf k} - {\mathbf k^\prime})~. 
\label{hel-vector-basis} \eea 
In other words, $A(k)$ represents the difference of the expectation
values of the positive and negative helicity field components. If $A$ does not
vanish, the left handed and right handed magnetic fields have
different strength.

We assume that both the symmetric and helical terms 
of the magnetic field power spectrum (\ref{spectrum}) can be approximated 
by a simple power law \cite{pogosian02}: 
\bea 
S(k) &=& \left\{ \begin{array}{ll}
S_0 ~k^{n_S}, & \mbox{ for } k<k_D \\ 
        0    &  \mbox{otherwise } 
\end{array} \right. \label{Sspec} \\
& \mbox{and} & \nonumber \\
 A(k) &=& \left\{ \begin{array}{ll}
   A_0 ~k^{n_A}, & \mbox{ for } k<k_D \\
        0  & \mbox{otherwise} 
\end{array} \right.
\label{Aspec} 
\eea 
where $S_0$, $A_0$ are the normalization constants, 
and $n_S$, $n_A$ the spectral indices of the symmetric 
and helical parts respectively. 
 
With (\ref{Sspec}, \ref{Aspec}), we can express the normalization constants 
$S_0$ and $A_0$ in terms of the averaged magnetic 
field energy density 
${B_\lambda}^2 \equiv \langle {\mathbf B}({\mathbf x}) 
\cdot {\mathbf B}({\mathbf x})\rangle|_\lambda$, 
and the absolute value of the averaged helicity 
${{\mathcal B}_\lambda}^2 \equiv 
\lambda|\langle {\mathbf B}({\mathbf x}) \cdot 
({\mathbf \nabla} \times {\mathbf B} ({\mathbf x}))\rangle ||_\lambda$ 
respectively, both smoothed over a sphere of comoving radius $\lambda$.
${\cal B}_\la$ measures the amplitude of helicity on the given comoving
scale $\lambda$.  
 
In order to calculate these quantities, we convolve the magnetic field 
and its helicity with a  3D-Gaussian filter function, so that 
$B_i\rightarrow B_i *f_\la$, where $\hat f_\la(k)=\exp(-\lambda^2 k^2/2)$. 
The mean-square values $B^2_\lambda$ and 
${\mathcal B}^2_\lambda$ 
are then given by the Fourier transform of the products of the corresponding 
spectra $S(k)$ and $k A(k)$ with the square of the filter function
$\hat f_\la$: 
\bea 
B^{2}_\lambda&=&\frac{1}{(2\pi)^3}\int d^3k\,S(k)\hat f_\la(k)^2 =
\frac{S_0}{(2\pi)^2}\frac{1}{\lambda^{n_S+3}} 
\Gamma\left(\frac{n_S+3}{2}\right)~, 
\label{B-mean} 
\\ 
 {\cal B}^2_\lambda&=&\frac{\lambda}{(2\pi)^3}\int d^3k\,k 
|A(k)| \hat f_\la(k)^2 = 
\frac{|A_0|} {(2\pi)^2}\frac{1}{\lambda^{n_A+3}}\Gamma 
\left(\frac{n_A+4}{2}\right)~. 
\label{hel-mean} 
\eea 
 
In order not to over-produce long range magnetic fields or helicity 
as $k \rightarrow 0$, we require  for the spectral indices $n_S >-3$ 
and $n_A > -4$ (for $n_A\le -3$ and $n_A\le -4$ the integrals (\ref{B-mean})
and (\ref{hel-mean}) diverge at small $k$). 
 
Using (\ref{B-mean}), (\ref{hel-mean}) and the definition of the 
magnetic field spectrum (\ref{spectrum}), we can rewrite 
expressions (\ref{B-vector-basis}) and (\ref{hel-vector-basis}) in 
the form (see also \cite{pogosian02}) 
\bea 
-\langle B^{+}{(\mathbf k)} B^{+}(-{\mathbf k^\prime}) + 
        B^{-}({\mathbf k}) B^{-}(-{\mathbf k^\prime})\rangle 
&=&(2\pi)^{5}\frac{\la^3B^2_\lambda}{\Gamma\left(\frac{n_S+3}{2}\right)} 
(\la k)^{n_S}\delta({\bk-\bk'})~, 
\label{Symm-2pt-fcn} 
\\ 
\langle B^{+}{(\mathbf k)} B^{+}(-{\mathbf k^\prime}) - 
 B^{-}({\mathbf k}) B^{-}(-{\mathbf k^\prime})\rangle 
&=&(2\pi)^{5}\frac{{\la^3\mathcal B}^2_\lambda 
}{\Gamma\left(\frac{n_A+4}{2} \right)} (\la k)^{n_A}\delta({\bk-\bk'}) 
\label{Hel-2pt-fcn}~, 
\eea 
for $k<k_D$ and $0$ for $k > k_D$. 
 
Using that 
$$ \lim_{\bk'\ra\bk}|\langle (\hat\bk\times \bB(\bk)) \cd \bB(-\bk')\rangle| 
\le \lim_{\bk'\ra\bk}\langle \bB(\bk)\cd \bB(-\bk')\rangle $$ 
we can conclude that 
\begin{equation}  \label{S>A} 
S(k) \geq |A(k)|~. 
\end{equation} Since $S(k) \propto \langle |\bB|^2\rangle$, it is clear 
that $S(k)\ge 0$. The reality condition requires $A_0$ to be real, but
it can be either positive or negative. For Eq.~(\ref{S>A}) to be
valid on very small values of $k$ requires  
\be n_A \geq n_S ~. \label{na>ns}\ee 
Applying Eq.~(\ref{S>A}) also close to the upper cutoff $k_D$, 
we have in addition 
\be 
 |A_0|  \le S_0k_D^{n_S-n_A}~. 
\label{limA} 
\ee 
In terms of the magnetic fields on scale $\la$ this gives roughly
\be \label{limBB}
 {\cal B}_\la^2 < B_\la^2(k_D\la)^{n_S-n_A} ~.
\ee
Usually the damping scale is much smaller than the physical scale of 
interest, $\la$ so that $\la k_D\gg 1$. Therefore, if $n_S-n_A \neq 
0$, the helical contribution is significantly suppressed on all scales 
$\la> \la_D= 1/k_D$. As we now show, this is always the case if the 
magnetic field is causally produced. 
 
Most mechanisms to produce magnetic fields with a helical component 
are causal. By this we mean that all correlations above a certain scale, 
usually some fraction of the Hubble scale at formation, have to vanish. If this 
is the case, causality implies an additional interesting constraint, 
which we now derive. For this we assume that  
the correlation functions $\langle B_i(\bx) B_j(\by)\rangle$ and 
 $\langle B_i(\bx)(\nabla\times \bB(\by))_j\rangle$ have to vanish 
 for $|\bx-\by|>R$ for some scale $R$. Hence they are functions with 
 compact support, which implies that their Fourier transforms, 
$P_{ij}S(k)$ and $\ep_{ijl}\hat k_lA(k)$ are analytic
 functions. Therefore, for 
 sufficiently small values of $k$ they can be approximated by power 
 laws as in Eqs.~(\ref{Sspec},\ref{Aspec}). Since $\hat k_j$ is not
 analytic but  $k\hat k_j$ is, this implies  
\be 
n_S \ge 2 \quad \mbox{ and }~~~ n_A \ge 1~, 
\ee 
where $n_S$ has to be an even integer while $n_A$ has to be an odd 
 integer. 
But since we need $n_A\ge  n_S$, this leaves us with 
\bea 
n_S &\ge& 2  \mbox{ , an even integer   and } \\ 
n_A &\ge& 3  \mbox{ , an odd integer}. 
\eea 
Causality together with the condition (\ref{S>A}) leads to 
an additional suppression of helical fields on large scales. 
Also ordinary causal magnetic fields cannot be white noise but are severely 
suppressed on large scale due to the non-analytic pre-factor $P_{ij}$ 
 in the power spectrum which is a simple consequence of the fact that 
 magnetic fields are divergence free $\nabla\cd\bB = 0$. This has 
 already been discussed in Refs.~\cite{durrer00,caprini02}. 
The causality constraint need not to be satisfied if the magnetic 
fields are generated before or during a period of inflation where the 
causal horizon diverges. For a detailed discussion of causality
see~\cite{Jcap}.

\section{Magnetic Source term for tensor metric perturbations} 
\label{sec:sources} 
 
The anisotropic stresses which act as source for metric perturbations 
are given by the magnetic field stress tensor \cite{jackson75} 
\be 
\tau_{ij}({\bf k}) = {1 \over (2\pi)^3} 
{1\over 4\pi}\int d^3 p \,[B_i({\bf p}) B_j^*({\bf p-k}) - 
{1\over 2}B_l({\bf p})B_l^*({\bf p-k})\de_{ij}]~. 
\label{taub} 
\ee 
Here we are interested in the generation of gravitational waves, and 
consequently we need to extract the transverse and traceless part of 
$\tau_{ij}$.  The form of a general projection to extract any 
mode (scalar, vector or tensor) 
from a generic tensorial perturbation can be found in \cite{dk98}. 
We make use of the tensor projector $T_{ijlm} = 
P_{il}P_{jm}-\frac{1}{2}P_{ij}P_{lm}$ (see also \cite{durrer00}). The 
tensor contribution to $\tau_{lm}$ is given by 
\be 
\Pi_{ij}=(P_{il}P_{jm}-\frac{1}{2}P_{ij}P_{lm})\tau_{lm}~. 
\label{tensor-proj} 
\ee 
Moreover, since the magnetic field is a stochastic variable, 
we need to calculate the two point correlation tensor of 
$\tau_{ij}({\mathbf k})$, which takes the form 
\be 
\langle \tau_{ij}({\mathbf k}) \tau^{*}_{lm}({\mathbf k^\prime}) 
\rangle = 
{1 \over (4 \pi)^2} {1 \over (2 \pi)^6} 
\int d^3p \int d^3q \,\langle B_i({\mathbf p})B_j({\bk - \bp}) 
B_l({- \bq}) 
B_m({\bq - \bk^\prime})\rangle + \cdots\de_{ij} + \cdots\de_{lm}~, 
\label{tau-2corr} 
\ee 
and we are not interested in terms proportional to 
$\de_{ij}$ and $\de_{lm}$ , which after being projected out 
will not contribute to the final result for the tensor perturbation 
$\langle \Pi_{ij}\Pi_{lm}\rangle$ (see appendix A. in \cite{mack02}). 
Before applying the tensor projection, %(\ref{tau-2corr}), 
we can simplify the right hand side of~(\ref{tau-2corr}) using Wick's 
theorem, expressing the four point correlators in terms of the two 
point ones, 
\bea 
\langle B_i({\mathbf k}_i)B_j({\mathbf k}_j) 
B_l({\mathbf k}_l)B_m({\mathbf k}_m)\rangle 
&=&\langle B_i({\mathbf k}_i)B_j({\mathbf k}_j)\rangle 
\langle B_l({\mathbf k}_l)B_m({\mathbf k}_m)\rangle 
\nonumber 
\\ 
&+ & 
\langle B_i({\mathbf k}_i)B_l({\mathbf k}_l)\rangle 
\langle B_j({\mathbf k}_j)B_m({\mathbf k}_m)\rangle 
\nonumber \\ 
& + & 
\langle B_i({\mathbf k}_i)B_m({\mathbf k}_m) 
\rangle\langle B_j({\mathbf k}_j)B_l({\mathbf k}_l)\rangle~. 
\label{Wick} 
\eea 
Since the two point correlation function given in Eq.~(\ref{spectrum}) is not 
symmetric, we are not allowed to change the order of indices $i,j,l,m$
inside an expectation value. 
With Eq.~(\ref{spectrum}) we can then compute the correlation function 
(\ref{tau-2corr}) which consists of a purely symmetric part 
proportional to $\int d^3p S(p)S(|{\mathbf k}-{\mathbf p}|)$~, 
a purely helical part  proportional to $\int d^3p A(p)A(|{\bk-\bp|})$~, 
and  mixed term, $i\int d^3p S(p)A(|{\mathbf k-\bp|})$
(the full expressions are given in Appendix~A, Eq.~(\ref{tau-appendix})). 
 The first two terms contribute to the 
symmetric part of the two point correlation function of 
the tensor source, while the two latter terms give rise to a helical
contribution. To express them we now introduce 
 the two point correlation function for the tensor 
source, which can be parameterized as 
\begin{equation} 
\langle\Pi_{ij}({\mathbf k})\Pi^{*}_{lm} 
({\mathbf k'})\rangle 
\equiv \frac{(2\pi)^3}{4} \left[ {\mathcal M}_{ijlm}f(k) + 
i{\cal A}_{ijlm} g(k) \right] \delta({\mathbf k-\bk'})~, 
\label{source-tensor} 
\end{equation} 
where the tensors ${\mathcal M}_{ijlm}$ and 
${\mathcal A}_{ijlm}$ are  given by 
\bea 
{\mathcal M}_{ijlm} 
&\equiv& P_{il}P_{jm}+P_{im}P_{jl}-P_{ij}P_{lm}~, 
\label{M_ijlm}\\ 
{\cal A}_{ijlm} &\equiv &{{\hat \bk}_q\over 2} 
   (P_{jm} \epsilon_{ilq} + P_{il} \epsilon_{jmq} + 
P_{im} \epsilon_{jlq} + P_{jl} \epsilon_{imq})~. 
\label{Aijlm} 
\eea 
Clearly, both ${\cal M}_{ijlm}$ and ${\cal A}_{ijlm}$ are symmetric 
in the first and second pair of indices. ${\cal M}_{ijlm}$ 
is also symmetric under the exchange of $ij$ with $lm$ while ${\cal 
  A}_{ijlm}$ is anti-symmetric under this permutation. 
We shall often use simple properties like 
\bea 
{\cal M}_{ijij} &=& 4~, \quad {\cal M}_{iilm} = {\cal M}_{ijll} = 0\\ 
P_{qi}{\mathcal M}_{ijlm} &= & {\mathcal M}_{qjlm} ~, \quad 
P_{qi}{\mathcal A}_{ijlm} =  {\mathcal A}_{qjlm}\\ 
{\mathcal M}_{ijlm}{\mathcal M}_{ijlm} &=& 
{\mathcal A}_{ijlm}{\mathcal A}_{ijlm} = 8 \\ 
{\cal A}_{ijlm}{\cal M}_{ijlm}  &=& 0 ~, \quad 
{\cal A}_{ijij} = {\cal A}_{iijl}= {\cal A}_{ijll}=0 ~. 
\eea 
According to Eq.~(\ref{tensor-proj}), we have now to act on 
$\langle \tau_{ab}({\mathbf k}) \tau^{*}_{cd} 
({\mathbf k^\prime}) \rangle$ with the tensor projector 
\be 
{\cal P}^{abcd}_{ijlm}(\hat\bk,\hat\bk') = (P_{ia}P_{jb} - 
  {1 \over 2} P_{ij}P_{ab})({\hat \bk})(P_{lc}P_{md} -
  {1 \over 2} P_{lm}P_{cd})({\hat \bk'})~. 
\label{tensor-projection} 
\ee 
In these calculations we don't need to care about the position (up
or down) of Latin
indices as they are always contracted by a Kronecker $\de$.
The symmetric and antisymmetric parts of Eq.~(\ref{source-tensor}) 
are invariant under the application of the projector 
(\ref{tensor-projection}), so that it is easy to separate the symmetric 
and helical parts of the source spectrum, $f(k)$ and $g(k)$: 
\bea 
(2\pi)^3\de(\bk -\bk')f(k) &=& {1 \over 2} {\mathcal M}_{abcd} 
\langle \tau_{ab}({\mathbf k}) \tau^{*}_{cd} 
({\mathbf k'}) \rangle 
\label{tensor-source-sym1} \\ 
(2\pi)^3\de(\bk -\bk')g(k) &=& {-i \over 2}{\cal A}_{abcd} 
\langle\tau_{ab}({\mathbf k})\tau^{*}_{cd} ({\mathbf k'}) 
\rangle \label{tensor-source-hel1}~. 
\eea 
Moreover, by applying the tensor  ${\mathcal M}_{ijlm}$ to Eq. 
(\ref{tau-appendix}) of Appendix A, we obtain (the first term of this 
has already been computed in Refs.~\cite{durrer00,caprini02,mack02}) 
\bea 
f(k)=\frac{2}{(4\pi)^5} \int d^3p\, \left[ 
\right. S(p)S(|{\mathbf k-\bp}|)(1+\gamma^2)(1+\beta^2) 
+4A(p)A(|{\mathbf k-\bp}|)(\gamma \beta) \left. \right]~, 
\label{tensor-source-2} 
\eea 
where $\gamma={\hat \bk}\cdot{\hat \bp}$ and $\beta={\hat \bk}\cdot (\widehat{ 
{\bf k-p}})$.  Note that the square of the  helical part 
of the magnetic field spectrum (\ref{spectrum}) contributes to the 
symmetric part of the source spectrum. This is not surprising, since 
the product of two quantities with odd parity has even parity. 
The antisymmetric part of the source spectrum is obtained by acting
with  ${\mathcal A}_{ijlm}$ on Eq.~(\ref{tau-appendix}) of Appendix A.
It is given by the mixed terms, 
\bea 
g(k)=  \frac{8}{(4\pi)^5} \int d^3p\, S(p)A(|{\mathbf 
k-\bp}|)(1+\gamma^2)\beta    \label{tensor-source-3}~. 
\eea 
 
We can also express the correlator~(\ref{source-tensor}) in terms of 
the  basis $e^{\pm }_{ij}$ introduced in~\cite{hu96}, 
\be 
{e}^{\pm }_{ij}= - \sqrt{\frac{3}{8}} ({\bf e}_1 \pm 
i{\bf e}_2)_i \times ({\bf e}_1 \pm i{\bf e}_2)_j~. 
\label{tensor-basis} 
\ee 
These form a basis of tensor perturbations,  satisfying the  
transverse-traceless condition $\delta_{ij} {e}^{\pm }_{ij}=0$, 
${\hat k}_i {e}^{\pm}_{ij}=0$ and  ${e}^{\pm }_{ij}{e}^{\mp 
}_{ij}=3/2$. Positive circularly polarized gravity waves are
proportional to $e^+_{ij}$, while negative  circularly polarized
gravity waves are given by the coefficient of $e^-_{ij}$. 
In this basis $\Pi_{ij}$ is expressed as 
\be 
\Pi_{ij}({\bf k}) \equiv e^{+ }_{ij} \Pi^+({\mathbf k}) + 
 e^{- }_{ij} \Pi^-({\mathbf k})~. 
\label{tensor-basis1} 
\ee 
We can rewrite $f(k)$ and $g(k)$ in terms of the components $\Pi^\pm$ as
\bea 
(2\pi)^3\de(\bk -\bk')~f(k) \equiv(2\pi)^3\de(\bk -\bk')~|\Pi(k)|^2&=& \frac{3}{2}~ 
\langle \Pi^{+}{(\mathbf k)} \Pi^{+*}({\mathbf k'}) + 
 \Pi^{-}({\mathbf k}) \Pi^{-*}({\mathbf k'})\rangle ~, 
\label{ft-tensor-basis}\\ 
(2\pi)^3\de(\bk -\bk')~g(k)&=& -\frac{3}{2}~ 
\langle \Pi^{+}{(\mathbf k}) \Pi^{+*}({\mathbf k'}) - 
 \Pi^{-}({\mathbf k}) \Pi^{-*}({\mathbf k'})\rangle~. 
\label{gt-tensor-basis} 
\eea 
Here we have used the form of $\cal M$ and $\cal A$ in this basis, 
\bean 
{\cal M}_{ijlm} &=& {4\over 3}\left[ e^{+ }_{ij}\otimes  e^{- }_{lm} 
   + e^{-}_{ij}\otimes  e^{+}_{lm}\right] \\ 
{\cal A}_{ijlm} &=& {4i\over 3}\left[ e^{+ }_{ij}\otimes  e^{- }_{lm} 
   - e^{-}_{ij}\otimes  e^{+}_{lm}\right] ~,
\eean 
and the simple properties of ${\cal M}_{ijlm}$ and ${\cal A}_{ijlm}$ 
mentioned above. One also has 
\be \label{Pieven} 
\langle \Pi^{+}{(\mathbf k)} \Pi^{-}({\mathbf k'})\rangle =
 \frac{(2\pi)^3}{3}~\de(\bk -\bk')~(f(k)+g(k))~.  
\label{traceless-PI} 
\ee
 
Similarly, defining the usual linear polarization basis 
\bea
e^T_{ij} &= &\left({\bf e}_1 \times{\bf e}_1 - {\bf e}_2 \times{\bf
  e}_2\right)_{ij}\nonumber \\
e^\times_{ij} &= &\left({\bf e}_1 \times{\bf e}_2 +{\bf e}_2 \times{\bf
  e}_1\right)_{ij}~, 
\label{tensor-basis2}
\eea
and the components of $\Pi$ with respect to this basis,
\bea
\Pi_{ij} &=& \Pi^Te^T_{ij} + \Pi^\times e^\times_{ij}~, \label{bas2}
\eea
we obtain also
\bea \label{Pieven2} 
\langle \Pi^{T}{(\mathbf k)} \Pi^{T*}({\mathbf k'}) + 
 \Pi^{\times}{(\mathbf k)} \Pi^{\times *}({\mathbf k'}) \rangle &=&
 (2\pi)^3\de(\bk -\bk')~f(k) \\
\langle \Pi^{\times}{(\mathbf k)} \Pi^{T*}({\mathbf k'}) - 
 \Pi^{T}{(\mathbf k)} \Pi^{\times*}({\mathbf k'}) \rangle  &=& 
i(2\pi)^3\de(\bk -\bk')~g(k)  ~.
\label{traceless-PI2} 
\eea

With Eqs.~(\ref{tensor-source-2}, \ref{tensor-source-3}), we find 
\bea 
%[\Pi^{-}({\bf k})+\Pi^{(2)}({\bf k})]^2 
 (2\pi)^3[f(k)+g(k)]=\frac{2}{(4\pi)^3}\int d^3p\, 
[S(p)(1+\gamma^2) + 2A(p) \gamma] \cdot [S(|{\mathbf 
k-\bp}|)(1+\beta^2) + 2A(|{\mathbf k-\bp}|)\beta]~. \label{ft-gt-total} 
\eea 
Let us introduce the tensor 
\be 
Q_{ij}({\bf k})\equiv\frac{1}{(4\pi)}[P_{ij}({\hat {\bf k}}) S(k) 
+ i \epsilon_{ijq} {\hat k}_q A(k)] 
\ee 
so that 
\be 
\frac{2}{(2\pi)^3} \frac{1}{4\pi} \langle B_i({\bf k}) B^*_j({\bf 
  k'})\rangle 
 = \de(\bk-\bk')Q_{ij}({\bf k})~; 
\label{q-tensor} 
\ee 
with $Q_{ij}(-{\bf k})=Q^*_{ij}({\bf k})$ one then finds 
\bea 
f(k)+g(k)%\equiv|\Pi_{tot}(k)|^2 
=\left [P_{ij}({\hat {\bf k}}) 
- i \epsilon_{ijq} {\hat k}_q\right] 
\left[ P_{lm}({\hat {\bf k}}) 
+ i \epsilon_{lmq^\prime} {\hat k}_{q^\prime}\right] 
\int ~d^3p~Q_{ij}({\bf p})Q^*_{lm}({\bf k-p}) ~. 
\label{q-relation} 
\eea 
%and then, taking $i=l$, $j=m$, we have: 
%\be 
%|\Pi_{tot}^{(T)}(k)|^2 = 4 
%\int ~d^3p~Q_{ij}({\bf p})Q^\star_{ij}({\bf p-k}) 
%\label{pi-tot-relation} 
%\ee 
%and thus we may express the $|\Pi_{tot}^{(T)}|^2$ in ${\bf x}$-space as: 
%\be 
%|\Pi_{tot}^{(T)}({\bf x})|^2 = 4 (2\pi)^3 
%Q_{ij}({\bf x})Q^\star_{ij}({\bf x}) 
%\nonumber 
%\\ 
%\ee 
 
Using Eqs. (\ref{Sspec}-\ref{hel-mean}), 
(\ref{tensor-source-2}) and (\ref{tensor-source-3}), it is possible to 
calculate $f(k)$ and $g(k)$. The details of the calculations are given in 
the Appendix A. The integrals cannot be computed analytically, but a good 
approximation gives, for $k<k_D$ (see also \cite{durrer00,mack02}): 
\bea 
f(k) &\simeq& {\mathcal A}_S \left((\la k_D)^{2n_S+3}+\frac{n_S}{n_S+3} 
(\la k)^{2n_S+3}\right) 
- {\mathcal A}_A \left((\la k_D)^{2n_A+3}+\frac{n_A-1}{n_A+4} 
  (\la k)^{2n_A+3}\right)    \label{T-S-Source} \\ 
g(k) & \simeq & {\mathcal C} \,(\la k_D)^{n_S+n_A+2}\, 
(\la k) \, \left[1 +  \frac{n_A-1}{n_S+3} \left(\frac 
{k}{k_D}\right)^{n_S+n_A+2}  \right] ~, \label{T-A-source} 
\eea 
 
where ${\mathcal A}_S$, ${\mathcal A}_A$ and ${\mathcal C}$ 
are positive  constants given in Eqs. (\ref{T-source-new}) to 
(\ref{T-source-hel-new}) of Appendix A. They depend on the 
 spectral indices $n_S$ and $n_A$ of the magnetic field 
 and on its amplitudes, which are given in terms of 
$B_\lambda^2$, ${\mathcal B}_\lambda^2$, and $\lambda$. 
 
Note that the contribution of magnetic field helicity to the symmetric part of 
the source, $f(k)$, is negative. But it is easy to check that 
Eq.~(\ref{S>A}) insures that it never dominates, hence $f\ge 0$. 
For $n_S, n_A >-3/2$, the two terms 
proportional to the upper cutoff $k_D^{2n_{S,A}+3}$ dominate in 
$f(k)$, which consequently depends only on the cutoff frequency and behaves 
like a white noise source \cite{durrer00}. For  $n_S<-3/2$ or also $n_A 
<-3/2$, the dominating terms go like $k^{2n_S+3}$ and 
$k^{2n_A+3}$ respectively. 
On the contrary, the antisymmetric source $g(k)$ never shows a white 
noise behavior. For $n_S+n_A>-2$ the dominant term is proportional to 
$k\,k_D^{n_S+n_A+2}$. For $n_S+n_A<-2$, $g(k)$ does not depend on the
upper cutoff, but is proportional to $k^{n_S+n_A+3}$. 
The singularities in the pre-factors ${\cal A}_S$, ${\cal A}_A$ and 
$\cal C$ which appear at $n_S=-3$ and $n_A=-4$ are the usual 
logarithmic singularities of scale invariant spectra. 
But as mentioned in Section~II the helical contribution must obey 
$n_A\ge n_S > -3$. The apparent singularities in the pre-factors at 
$n_{S,A}=-3/2$ and at $n_S+n_A=-2$ are removable when multiplied with 
the $k$-dependent parts as in Eqs.~(\ref{T-S-Source}) and (\ref{T-A-source}). 
In the integrals over $k$ which we shall perform to calculate the
$C_{\ell}$'s we only take into account the dominant terms. 

If the magnetic field is causal, we expect $n_S=2$ and $n_A=3$, so 
that 
\bea 
f(k) &\simeq&  {\mathcal A}_S(k_D\la)^7 -{\cal A}_A(k_D\la)^9 \\ 
g(k) &\simeq& {\cal C}k\la (k_D\la)^7~. 
\eea 
Comparing the limit given in Eq.~(\ref{limA}) with the expressions for 
$ {\mathcal A}_S$ and $ {\mathcal A}_A$ derived in the Appendix~A, it 
is easy to see that $f$ always remains positive. 
 
The analysis of the evolution of a non-helical magnetic field 
interacting with the primordial plasma, and the derivation of the
appropriate damping scale $k_D$, has been discussed in 
Refs.~\cite{jedamzik98} and~\cite{subramanian98a}, 
where the authors considered a magnetic field with a tangled component 
superimposed on a homogeneous 
field. We assume that the latter can be obtained 
by smoothing our stochastic field on a scale which is 
larger than the damping scale (for details, see
~\cite{durrer00,caprini02}). 
The damping scale for the tensor mode is obtained 
taking into account that the source of gravitational radiation after 
equality becomes sub-dominant so that the relevant tensor 
damping scale is the Alfv\'en wave damping scale from the time 
of the creation of the magnetic field up to equality 
\cite{caprini02}. Since we are interested here in the 
imprint of the magnetic field on the CMB, we need not to care 
about the time evolution of the damping scale, the relevant 
scales for the CMB tensor anisotropies being those which are 
greater or equal to the horizon at equality. Therefore, the 
relevant cutoff scale is given by the Alfv\'en 
wave damping scale at equality 
$k^{-1}_D\simeq v_A 
l_\gamma(T_{\text{eq}})$, where 
$l_\gamma(T_{\text{eq}}) \approx 0.35\,{\rm Mpc}$ 
is the comoving diffusion length of photons at equality (here we
have used that $l^{\rm phys}_\gamma(T)\simeq
10^{22}\text{cm}(T/T_{\text{dec}})^{-3}$, from \cite{subramanian98a},
as well as $z_{\text{eq}}\simeq 3454$ and $z_{\text{dec}}=1088$ from the WMAP
results \cite{spergel}). The Alfv\'en speed is at most of order
$10^{-3}$, so that the damping scale is on the order of kpc or smaller.

Even if considering an helical component in the magnetic field, we set 
all the power to zero on scales smaller than $k_D^{-1}$. This is not
really correct since simulations show~\cite{mark} that the
spectrum simply decays like a power law with index of the order of
$-4$ on small scales, $k>k_D$.
However, as we shall see, for $n_{S,A}<-3/2$ the induced $C_\ell$'s are
dominated by   the contribution at the largest scales, $k_D^{-1}$, for the
kinks, $n_{S,A}\sim -4$ part of the spectrum. Therefore, we do not loose
much by neglecting the contribution from the scales smaller than $k_D^{-1}$.

\section{Magnetic Field induced tensor metric perturbations} 
 
A stochastic magnetic field can act as a source for Einstein's 
equations and hence generate gravitational waves, see for example 
\cite{durrer00,mack02,caprini02}. The tensor modes are the 
simplest case of metric perturbations, and in the transverse and 
traceless gauge they are fully described by the tensor 
 $h_{ij}({\mathbf x},\eta)$, satisfying 
\be 
h_{ij}=h_{ji}, ~~~~~h_{ii}=0,~~~~~h_{ij}{\hat k}^j=0~. 
\ee 
The linear evolution equation for gravitational waves is 
\be 
\ddot{h}_{ij}({\mathbf k},\eta)+2\frac{\dot{a}}{a} 
\dot{h}_{ij}({\mathbf k},\eta)+k^2h_{ij}({\mathbf k},\eta) 
=\frac{8\pi G}{a^2(\eta)}\Pi_{ij}({\mathbf k}), 
\label{T-Einstein} 
\ee 
where $\Pi_{ij}({\mathbf k})$ is the source tensor given in 
(\ref{tensor-proj}), and we have multiplied in the time dependence
$a^{-2}(\eta)$, which comes from the fact that the magnetic field is
frozen in the plasma. Therefore, $\Pi_{ij}({\mathbf k},\eta)$ is a
coherent source, in the sense that each mode undergoes the same time
evolution \cite{caprini02}.  
We neglect other possible anisotropic stresses of the plasma 
(collisionless hot dark matter particles or massless neutrinos have
anisotropic stresses which do source gravitational waves, but this
effect is very small \cite{durrer98b}).  
 
We want to compute the induced CMB 
anisotropies and polarization (see Section~V), which 
can be expressed in terms of the two-point correlation spectrum 
$\langle\dot h_{ij}({\bf k}) \dot h_{lm}({\bf k}^\prime)\rangle$,  
taking the form~\cite{durrer00,caprini02}: 
\be 
\langle\dot h_{ij}({\bf k}, \eta) \dot h^*_{lm}({\bf k}^\prime ,\eta )\rangle 
= \frac{(2\pi)^3}{4} \left[ {\mathcal M}_{ijlm} H(k,\eta) + i \A_{ijlm}\HH(k,\eta) 
\right] \delta({\bf k} - {\bf k^\prime})  ~. 
\label{gw} 
\ee 
Here $ (2\pi)^3H(k,\eta)\delta({\bk-\bk'}) = 
\langle\dot h_{ij}({\bf k}) \dot h^*_{ij}({\bf k^\prime})\rangle$ 
 is the usual isotropic part of 
the gravitational wave spectrum which is sourced by 
$f(k)$, and ${\mathcal H}(k, \eta)$ describes the 
helical part, sourced by $g(k)$. 
  
The perturbation tensor $h_{ij}$ can also be expressed in terms of the 
basis ${e}^{\pm}_{ij}$ defined in Eq.~(\ref{tensor-basis}): 
\be 
h_{ij}({\bf k}, \eta)=h^{+}({\bf k},\eta) {e}^{+}_{ij} + 
h^{-}({\bf k},\eta) {e}^{-}_{ij}~. 
\label{gw-def} 
\ee 
Just like for the anisotropic stress power spectra, we now find that 
\bea 
 (2\pi)^3\de(\bk -\bk')~H(k, \eta) &\equiv& 
\frac{3}{2} \langle 
\dot h^{+}({\bf k},\eta) \dot h^{+*}({\bf k'},\eta) 
+ \dot h^{-}({\bf k}, \eta) \dot h^{-*}({\bf k'},\eta)\rangle~, 
\label{gw-sym-basis} 
\\ 
 (2\pi)^3\de(\bk -\bk')~{\mathcal H}(k, \eta) &\equiv& -{3\over 2} 
 \langle \dot h^{+}({\bf k},\eta) \dot h^{+*}({\bf k'}, \eta) 
-\dot h^{-}({\bf k},\eta) \dot h^{-*}({\bf k'},\eta)\rangle~. 
\label{gw-cross-basis} 
\eea 
In terms of $h^T$ and $h^\times$, defined like in Eq.~(\ref{tensor-basis2}),
$\cal H$ parameterizes the correlation between $h^T$ and $h^\times$,
\be 
\langle \dot h^{\times}{(\mathbf k)} \dot h^{T*}({\mathbf k'}) - 
 \dot h^{T}{(\mathbf k)} \dot h^{\times*}({\mathbf k'}) \rangle  = 
i (2\pi)^3\de(\bk -\bk')~{\cal H}(k)~.
\ee
The evolution equation for the components $h^{\pm}(\bk,\eta)$ is simply 
\be 
{\ddot{h}}^{\pm}({\mathbf k},\eta)+2\frac{\dot{a}}{a} 
{\dot{h}}^{\pm}({\mathbf k},\eta)+k^2h^{\pm }({\mathbf k},\eta) 
=\frac{8\pi G}{a^2(\eta)}\Pi^{\pm }({\mathbf k})~. 
\label{T-Einstein-basis} 
\ee 
We need to determine the 
functions $\dot{h}^{\pm}({\bf k},\eta)$ (see
Eq.~(\ref{T-temp-int-soln-1}) below). 
An approximate solution to the above differential equation can be found in 
\cite{durrer00} or \cite{caprini02}. The important point is that 
because of the rapid falloff of the magnetic field source in 
the matter dominated era, 
perturbations created after equality ($\eta_{\text{eq}}$) are sub-dominant, 
so that one obtains, for the dominant contribution at $\eta>\eta_{\text{eq}}$: 
\be 
\dot{h}^{\pm }({\bf k},\eta) \simeq {16\pi G\over 
  H_0^2\Om_r} 
\ln\left(\frac{z_{\text{in}}}{z_{\text{eq}}}\right) 
\Pi^{\pm }({\bf k})\, 
\frac{j_2(k\eta)}{\eta} ~, 
%\simeq  \frac{1}{\rho_c\Om_r} 
%\ln\left(\frac{z_{\text{in}}}{z_{\text{eq}}}\right) \Pi^{\pm 2}({\bf k})\, 
%\frac{j_2(k\eta)}{\eta}~, 
\label{T-rms-metric} 
\ee 
where $\Omega_r$ is the radiation density parameter today and 
$z_{\text{in,eq}}$ correspond 
to the redshifts at the moment of creation of the magnetic field and 
at matter radiation equality respectively. The function $j_2$ is the 
spherical Bessel function~\cite{abramowitz72}. The term 
$\ln(z_{\rm in}/z_{\rm eq})$ accounts for the logarithmic build up of
gravity waves from $z_{\rm in}$ to $z_{\rm eq}$.
For the spectra~(\ref{gw-sym-basis}) and (\ref{gw-cross-basis}) we 
then obtain 
\bea 
H(k,\eta) &\simeq&  \left[{16\pi G\over H_0^2\Om_{r}} 
  \ln\left(\frac{z_{\text{in}}}{z_{\text{eq}}}\right) 
  \frac{j_2(k\eta)}{\eta}\right]^2f(k) ~,\\ 
\HH (k,\eta)&\simeq&  \left[{16\pi G\over H_0^2\Om_r} 
  \ln\left(\frac{z_{\text{in}}}{z_{\text{eq}}}\right) 
  \frac{j_2(k\eta)}{\eta}\right]^2 g(k) ~. \label{violwave}
\eea 
 
The gravity wave power spectra $H/\rho_r$ and ${\cal H}/\rho_r$ are
constant on large scales, $k\eta\ll 1$ and decay and oscillate inside
the horizon.

Our first result is that a helical magnetic field induced a parity odd
gravity wave component. From Eq.~(\ref{T-Einstein-basis}) it is clear,
that such a component is introduced whenever there are parity odd
anisotropic stresses. It could {\em in principle} also
be detected directly, via gravity wave background detections
experiments. We do not discuss this very hypothetical idea any
further, but calculate the effect of such a component on CMB
anisotropies and polarization.

%The presence of non-zero helical part ${\mathcal H}(k)$ is 
%reflecting the fact of possibility of gravitational waves mixing. 
%The physical explanation of non-zero  ${\mathcal H}(k)$ becomes clear 
%using linear polarization tensor basis, in which 
%${\mathcal H}(k) \sim H^{(+)}({\bf k})H^{(\times)}({\bf k})$. 

\section{CMB fluctuations} 
\label{CMB-fluctuations} 
 
Magnetic fields in the universe lead to all types of  metric 
perturbations (scalar, vector and tensor, 
for more details see \cite{grasso01}). 
In~\cite{mack02} it is shown that 
vector and tensor perturbations from magnetic fields induce  CMB 
anisotropies of the same order of magnitude. 
In this paper we 
estimate CMB fluctuations due to gravitational waves induced 
by a stochastic magnetic field, the spectrum of which contains an helicity 
component, $A(k)\neq 0$. 
Since the CMB signature of chaotic  magnetic fields with only an isotropic 
spectrum is given in detail in Refs.~\cite{durrer00,mack02}, 
here we concentrate on the effects from 
the helical part of the magnetic field spectrum, and we will discuss 
the corrections which it induces to the previous results. 
 
To compute the CMB fluctuation power spectra we use 
the total angular momentum method introduced by Hu 
and White \cite{hu96}. By combining intrinsic angular 
structure with  the spatial dependence of plane-waves, Hu and White 
obtained integral solutions for all kind 
of perturbations. 
% given by eqs.(74),(76)-(77) in 
The angular power spectrum of CMB fluctuations can then be expressed 
as~\cite{hu96} 
\be 
C^{{\mathcal X},{\mathcal X}^\prime}_{\ell} 
={2 \over \pi} 
\int dk\,k^2 \sum_{m=-2}^{+2} 
{{\mathcal X}_{(m)\ell}(k,\eta_0) \over 2\ell+1} 
{{{\mathcal X}^{\prime *}}_{(m)\ell}(k,\eta_0) \over 2\ell+1}~, 
\label{C-power-spectrum} 
\ee 
where $\mathcal X$ takes the values of $\Theta$, temperature 
fluctuation, $E$, polarization with positive parity, and $B$, 
polarization with negative parity, for each 
perturbation mode. The index $m$ indicates the spin, and for tensor modes 
$m=\pm 2$. Since we only consider tensor modes in this paper, we suppress 
the index $2$ and just denote the two states by $+$ and $-$ in 
what follows. 
The description given in Ref.~\cite{mack02} applies the total angular 
momentum method to parity even magnetic field spectra: in this case, 
according to parity conservation the sum over $\pm $ can be 
replaced by a factor $2$. In our case instead, we always need to
sum over both states. 
 
From the form of $f(k)$, the parity even CMB fluctuation correlators 
can be expressed as: 
\be 
C^{{\mathcal X},{\mathcal X}^\prime}_{\ell}= 
C^{{\mathcal X},{\mathcal X}^\prime}_{(S)\ell}- 
C^{{\mathcal X},{\mathcal X}^\prime}_{(A)\ell}~, 
\label{decomposition_Cl} 
\ee 
where $C^{{\mathcal X},{\mathcal X}^\prime}_{(A)\ell}$ is the power spectrum 
induced by the purely helical part of the source term, proportional to 
 $A(p)A(|{\bf k- p}|)$. The contribution of this helical part
to the parity even  CMB power spectra is always negative, but, as we 
shall see, the condition (\ref{S>A}) insures that $C^{{\mathcal X},{\mathcal 
    X}}_{(A)\ell} < C^{{\mathcal X},{\mathcal X}}_{(S)\ell}$ 
so that the power spectra do not become negative. 
 
The new effect is that the helical part of the magnetic field now also 
induces  parity odd CMB correlators, $C_\ell^{\Theta B}$ and $C_\ell^{EB}$ 
(see also~\cite{pogosian02}). These are expressed in terms of the 
helical magnetic source $g(k)$ which is proportional to the
convolution of  $A(k)$ with   $S(k)$ (see Eq.~(\ref{tensor-source-3})). 
 
We now derive the CMB fluctuations $\Th^\pm_{\ell}(\eta_0,k)$, 
$E^\pm_{\ell}(\eta_0,k)$, $B^\pm_{\ell}(\eta_0,k)$ and then perform the 
integral (\ref{C-power-spectrum}). Rather than a numerical study, we 
present analytical approximations for our results. These are not very 
accurate, but allow a discussion of the dependence of the correlators
on $n_S$ and $n_A$. We will also be able to determine the spectral 
index of the CMB correlators (dependence on $\ell$) as a function of
$n_S$ and $n_A$. At the present stage, we think this scaling
information is more interesting than accurate numerical results. These
can than follow for specific, interesting values of the spectral
indices in future work. For a magnetic field with no helical component, this 
program has been carried out in Ref.~\cite{mack02}, and we shall just 
refer to their results but not re-derive them here. 
 
Below, we shall always work in the approximation of `instant 
recombination'. Moreover, in our approximations we didn't take into
account the decay of gravity waves for modes which entered the horizon
before decoupling. Our results therefore will be reasonable 
approximations (within a factor of two or so) only for $\ell\lsim 60$,
where the tensor CMB signal is  largest. Even 
though, this may seem poor accuracy, here  we only want to obtain
estimates of the correct order of magnitude of this anyway small
effect. This will enable use to judge for which cases a more involved
numerical study is justified. 
 
\subsection{CMB temperature anisotropies} 
 
Within the instant recombination approximation, gravitational waves 
simply cause CMB photons to propagate along perturbed geodesics 
from the last scattering surface to us. The induced CMB temperature 
anisotropies are given by~\cite{durrer94} 
\begin{equation} 
\Theta(\eta_0,{\mathbf k},\hat{{\mathbf n}}) 
\simeq  \int^{\eta_0}_{\eta_{\text{dec}}}d\eta\, 
         \exp(-i(\eta_0-\eta)\bk \cdot \bn) 
         \dot{h}_{ij}({\mathbf k},\eta){\hat n}_i{\hat n}_j~. 
\label{T-CMB-geodesics} 
\end{equation} 
In the total angular momentum formalism this becomes 
\be 
\frac{\Theta^{\pm}_{\ell}(\bk,\eta_0)}{2\ell+1}= -{4\over 3} 
\int^{\eta_0}_{\eta_{\text{dec}}} d\eta\, {\dot{h}}^{\pm }(\bk,\eta) 
j^{\pm }_{\ell}[k(\eta_0-\eta)]~, 
\label{T-temp-int-soln-1} 
\ee 
where $j^{\pm }_{\ell}$ are the tensor 
temperature radial functions of the two different parities, both given 
by~\cite{hu96}
\begin{equation} 
j^{\pm}_{\ell}(x)=\sqrt{\frac{3}{8}\frac{(\ell+2)!}{(\ell-2)!}}\,
\frac{j_{\ell}(x)}{x^2}~. 
\label{T-temp-radial-fcn} 
\end{equation} 
The somewhat unusual factor $4/3$ comes from the fact that this formula 
takes into account polarization, while Eq.~(\ref{T-CMB-geodesics}) does 
not. A detailed derivation can be found in Ref.~\cite{hu96}. 
 
Using the solution (\ref{T-rms-metric}) for 
$\dot{h}^{\pm }(\bk,\eta)$, we obtain 
\bea
\frac{\Theta^{\pm }_{\ell}(\bk,\eta_0)}{2\ell+1} & \simeq &
-\sqrt{\frac{3}{8}\frac{(\ell+2)!}{(\ell-2)!}}
\left[{8\over \rho_c\Om_r}
\ln\left(\frac{z_{\text{in}}}{z_{\text{eq}}}\right)\right]
\Pi^{\pm}(\bk)\int^{x_0}_{x_{\text{dec}}} dx\,
\frac{j_2(x)}{x}\frac{j_{\ell}(x_0-x)}{(x_0-x)^2}
\nonumber \\
 & \simeq & {-2\over \rho_c\Om_r}
\ln\left(\frac{z_{\text{in}}}{z_{\text{eq}}}\right)
\Pi^{\pm}(\bk)\frac{J_{\ell+3}(x_0)}{x^3_0}\,\ell^{5/2}
\label{T-temp-int-soln-2}
\eea
where we have set $x\equiv k\eta$ and $x_0\equiv k\eta_0$. For the 
 second $\simeq$ sign we have used the 
 approximation~(\ref{approx1}) given in Appendix~B for the integral
 over $x$. This approximation is valid only for $x_{\rm dec}
 =k\eta_{\rm dec}\lsim 1$. 
%The following integrals over $k$ are therefore valid
% only if they are dominated by values of $k$ with $k\eta_{dec}\lsim
% 1$. Generically this is the case for $\ell\lsim 60$. Therefore, our
% approximations hold for $10\lsim \ell \lsim 60$.
 
The general expression~(\ref{C-power-spectrum}) for the temperature 
anisotropy power spectrum now gives 
\be  \label{T1}
C^{\Theta\Theta}_{\ell} \simeq {16\over 3\pi} \left[{1 \over
\rho_c\Om_r}\ln\left(\frac{z_{\text{in}}}{z_{\text{eq}}}\right)\right]^2
\frac{\ell^5}{\eta_0^{3}}\int_0^{k_D\eta_0}dx_0 \, 
    {J^2_{\ell+3}(x_0)\over x_0^4}f\left(\frac{x_0}{\eta_0}\right) ~. 
\ee 
 
A good approximation for the function $f(k)$ is given in Appendix~A, 
Eq.~(\ref{T-source-appendix}). The first term of
(\ref{T-source-appendix}) comes entirely from the 
non-helical component $B_\la$, and has already been determined in 
Refs.~(\cite{durrer00,mack02}); the second term comes instead from the 
helical component, and its influence on the $C_{\ell}$ is new. 
We denote it by $C^{\Theta\Theta}_{(A)\ell}$. 
Then, splitting the induced temperature anisotropy power spectrum as 
\be 
C^{\Theta\Theta}_{\ell} = C^{\Theta\Theta}_{(S)\ell} - 
      C^{\Theta\Theta}_{(A)\ell}~, 
\ee 
we obtain (now $x_0$ is renamed $x$)
\be 
C^{\Theta\Theta}_{(A)\ell} 
\simeq {2^9\pi\over 9}{ \big[{\Om_A \over \Om_r}
\ln \big(\frac{z_{\text{in}}}{z_{\text{eq}}}\big)\big]^2
\over (2n_A+3)\Gamma^2\left(\frac{n_A+4}{2}\right)}  
\ell^5 
\left({1\over \eta_0 k_D}\right)^3 
\int_0^{x_D}dx \,  {J^2_{\ell+3}(x)\over x^4}
\left[1+\frac{n_A-1}{n_A+4} \,\left({x\over x_D}\right)^{2n_A+3}\right]~, 
\label{T-temp-A-power-spectrum-1} 
\ee 
where we have set $x_D=k_D\eta_0$. 
We have introduced the `helicity density parameter' $\Om_A$ defined by 
\be \label{densparA} 
\Om_A \equiv {\BB_\la^2\over 8\pi\rho_c}(k_D\la)^{n_A+3} \simeq
    {1\over \rho_c} \int_0^{k_D} {dk\over k}{d\rho_{\cal B}(k)
      \over d\log k}  
   \simeq {\BB_{k_D}^2\over 8\pi\rho_c}~, 
\ee 
%***************************************************************************\\ 
%MORE PRECISELY WE COULD DEFINE $\Om_A' \equiv {\BB_\la^2\over
%  8\pi\rho_c}(k_D\la)^{n_A+3} (2\pi)^2/((n_A+3) \Gamma({n_A+4 \over
%  2}))$. \\  DO YOU THINK WE NEED THIS?\\  
%***************************************************************************\\
and analogously we will use 
\be \label{densparS} 
\Om_S \equiv {B_\la^2\over 8\pi\rho_c}(k_D\la)^{n_S+3} \simeq
    {1\over \rho_c} \int_0^{k_D} {dk\over k}{d\rho_B(k)
      \over d\log k}  
   \simeq {B_{k_D}^2\over 8\pi\rho_c}~, 
\ee 
where we have introduced $\BB_{k_D}^2=\BB_\la^2(k_D\la)^{n_A+3}$, the
field strength at the cutoff scale $1/k_D$, and correspondingly for $B_{k_D}$.
With these definitions the results will be expressed entirely in terms
of physical quantities and the reference scale $\la$ does no longer enter.

Remember also that $(2\pi)^{4}({\mathcal B}_\lambda^2 \lambda^{n_A+3})^2 
/\Gamma^2\left(\frac{n_A+4}{2}\right) =  |A_0|^2$, where $|A_0|$ is 
the normalization of the helical component of the magnetic power spectrum 
(\ref{Aspec}).  The integral (\ref{T1}) is
dominated at $x_0\simeq
\ell$. With $x_0/x_{\rm dec} =\eta_0/\eta_{\rm dec} \simeq 60$, this
means that our approximation is valid for $\ell\lsim 60$.

If $n_A>-3/2$, the first term in the square bracket in
Eq. (\ref{T-temp-A-power-spectrum-1}) 
dominates. Since the integral converges and is 
maximal around $k \simeq \ell/\eta_0 \ll k_D$, we can replace it by 
the integral to infinity and use Eq.~(\ref{eq:GR-6.574.2}) of Appendix~B. 
This gives 
\bea 
\ell^2 C^{\Theta\Theta}_{(A)\ell} \simeq {2^{10} \over 27}
{\big[{\Om_A\over \Om_r} \ln\big(\frac{z_{\text{in}}}{z_{\text{eq}}}\big)
\big]^2 \over (2n_A+3)\Gamma^2\left(\frac{n_A+4}{2}\right)} 
\left({\ell\over k_D\eta_0}\right)^3 \label{T-temp-power-spectrum-A} 
\\
& & \qquad \mbox{ for } n_A>-3/2~. \nonumber 
\eea 
The temperature 
power spectrum has the well known behavior of $C_\ell$'s induced by 
white noise gravity waves, $C_\ell \propto \ell$. 
 
If $n_A<-3/2$, the second term in the square bracket of 
Eq.~(\ref{T-temp-A-power-spectrum-1}) dominates, and we find 
\bea 
\ell^2C^{\Theta\Theta}_{(A)\ell} \simeq 
\frac{2^8\sqrt{\pi}}{9}
{\big[{\Om_A\over \Om_r} \ln\big(\frac{z_{\text{in}}}{z_{\text{eq}}}\big)
\big]^2 \over (2n_A+3)\Gamma^2\left(\frac{n_A+4}{2}\right)}
{\Gamma\left(\frac{1}{2}-n_A\right) \over \Gamma(1-n_A)} 
{n_A-1 \over n_A+4}    
\left({\ell\over k_D\eta_0}\right)^{2n_A+6}\label{T-temp-power-spectrum-B} \\ 
& &\qquad \mbox{ for }  -3<n_A<-3/2~. \nonumber
\eea 
 
Like for the symmetric contribution given in 
Refs.~\cite{durrer00,mack02}, we get a scale-invariant spectrum for 
$n_A=-3$.  The expressions for $\ell^2C^{\Theta\Theta}_{(S)\ell}$ 
are obtained from those given above upon replacing $\Om_A$ by 
$\Om_S$, $n_A$ by $n_S$ and $\Gamma^2\left(\frac{n_A+4}{2}\right)$ by 
$\Gamma^2\left(\frac{n_S+3}{2}\right)$. For $-3< n_S<-3/2$, one also has 
to replace the factor $(n_A-1)/(n_A+4)$ by $n_S/(n_S+3)$. 
We do not repeat these formulas here since they can be found in 
Ref.~\cite{mack02} (up to some factors of order unity which are of no 
relevance for this discussion). 
 
This is in principle the final result for temperature anisotropies. 
Let us check that $C^{\Theta\Theta}_{(A)\ell}$ is indeed never larger 
than $C^{\Theta\Theta}_{(S)\ell}$ so that 
\[ C^{\Theta\Theta}_{\ell} = C^{\Theta\Theta}_{(S)\ell} - 
   C^{\Theta\Theta}_{(A)\ell} \ge 0~.\] 
 
We first consider $n_A\ge n_S>-3/2$. Then 
\be 
 {C^{\Theta\Theta}_{(A)\ell} \over C^{\Theta\Theta}_{(S)\ell}} 
   =  {\BB_\la^4\,\Ga^2({n_S+3\over 2})\,(2n_S+3)\,(k_D\la)^{2(n_A-n_S)} \over 
          B_\la^4\,\Ga^2({n_A+4\over 2})\,(2n_A+3)} 
= {|A_0|^2\over S_0^2\,k_D^{2(n_S-n_A)}} \frac{2n_S+3}{2n_A+3} \le 
1~.
\ee 
In the first equality we have inserted the definitions of $\Om_A$ and 
$\Om_S$ and the last inequality comes from Eqs.~(\ref{limA}) 
and~(\ref{na>ns}). 
If instead  $n_S\le n_A < -3/2$, we find 
\be 
 {C^{\Theta\Theta}_{(A)\ell} \over C^{\Theta\Theta}_{(S)\ell}} 
%  & = & \left[ 2^{2(n_A-n_S)}{\Ga(1-2n_A)\Ga^2(1-n_S)(n_A-1)(2n_S+3)(n_S+3) 
%   \over \Ga(1-2n_S)\Ga^2(1-n_A)n_S(2n_S+3)(n_A+4)} \right] 
%{\BB_\la^4\Ga^2({n_S+3\over 2})\la^{2(n_A-n_S)} \over 
% B_\la^4\Ga^2({n_A+4\over 2})} 
%\left(\ell/\eta_0\right)^{2(n_A-n_S)} \nonumber\\ 
= N(n_A,n_S){|A_0|^2\over S_0^2\,k_D^{2(n_S-n_A)}} 
                \left({\ell\over k_D\eta_0}\right)^{2(n_A-n_S)}~, 
\ee 
where $N(n_A,n_S)$ is a function of the spectral indices $n_S$ and 
$n_A$. It is of 
order unity in the allowed range, $-3<n_A\le n_S<-3/2$. Now $k_D\eta_0\gg 
\ell$ for all values of $\ell$ for which our result applies. Hence 
again 
\be 
{C^{\Theta\Theta}_{(A)\ell} \over C^{\Theta\Theta}_{(S)\ell}} 
\le 1~. 
\ee 
 
Finally, we consider the case $-3<n_S<-3/2<n_A$, so that we have to 
apply the result~(\ref{T-temp-power-spectrum-A}) for 
$C^{\Theta\Theta}_{(A)\ell}$ and 
(\ref{T-temp-power-spectrum-B}) with the mentioned modifications 
for $C^{\Theta\Theta}_{(S)\ell}$. 
A short calculation gives 
\be 
{C^{\Theta\Theta}_{(A)\ell} \over C^{\Theta\Theta}_{(S)\ell}} 
 \simeq {|A_0|^2\over S_0^2\,k_D^{2(n_S-n_A)}} \left({k_D\eta_0 \over 
     \ell}\right)^{2n_S+3} \le 1~, 
\ee 
since the first factor is less than one due to Eq.~(\ref{limA}) and 
$k_D\gg \ell/\eta_0$ with $n_S<-3/2$. 
 
Clearly, the helical component is maximal for $n_A \simeq n_S$, where we may 
have $|A_0|\simeq S_0$.

\subsection{The induced CMB polarization} 
\label{sub-CMBpolarization}
 
Tensor perturbations induce both $E$ polarization with 
positive parity, and $B$ polarization with negative parity. 
CMB polarization induced by gravity waves has been studied for example in 
Refs.~\cite{hu96,zaldarriaga97,kamionkowski97b}, while the 
contribution from a magnetic field has been discussed 
in~\cite{kosowsky96,mack02}. Our 
aim is to estimate the effect on the polarization signal from the 
helical component of the magnetic field. Like for the temperature 
anisotropies, we use 
the angular momentum method developed in Ref.~\cite{hu96}. 
 
\subsubsection{$E$ type polarization} 
 
The integral solution for $E$ type polarization from gravity waves is 
given in~\cite{hu96}. Again, we will work in the `instant 
recombination' approximation. The 
order of magnitude of our result is still reasonable for
$\ell \lsim 60$, since in this case also we restrict ourselves to the
evaluation of the super-horizon scales spectrum. 
In our approximation we have 
\begin{equation} 
\frac{E^{\pm }_\ell(\bk,\eta_0)}{2\ell+1}= \sqrt{2\over 
  3}\int^{\eta_0}_{\eta_{\text{dec}}} d\eta\, 
{\dot h}^{\pm}(\bk, \eta)\epsilon^{\pm }_\ell[k(\eta_0-\eta)]~, 
\label{T-E-int-soln-1} 
\end{equation} 
here 
\begin{equation} 
\epsilon^{\pm }_{\ell}(x)=\frac{1}{4}\left[-j_\ell(x)+j''_\ell(x) 
+2\frac{j_\ell(x)}{x^2}+4\frac{j'_\ell(x)}{x}\right] 
\simeq \frac{1}{4}\left[{\ell^2\over x^2}j_\ell -2j_\ell(x)\right] 
\quad \mbox{ for } \ell\gg 1 
\label{T-E-radial-fcn} 
\end{equation} 
is the E-type polarization radial function for the tensor mode 
\cite{hu96}, and for the last equality we have used the recurrence relations 
for spherical Bessel functions 
(\ref{eq:AS-10.1.21}, \ref{eq:AS-10.1.22}). 
 
We now use our solution~(\ref{T-rms-metric}) to express 
$\dot{h}^{\pm}(\bk,\eta)$ in 
terms of $\Pi^{\pm}(\bk)$. With this, Eq.~(\ref{T-E-int-soln-1}) becomes 
\begin{eqnarray} 
\frac{E^{\pm}_\ell(\bk,\eta_0)}{2\ell+1} 
&\simeq& \sqrt{\frac{3}{2}} 
\left[{1\over \rho_c\Om_r} 
       \ln\left(\frac{z_{\text{in}}}{z_{\text{eq}}}\right)\right] 
\Pi^{\pm}(\bk) \int^{x_0}_{x_{\text{dec}}}dx\,\frac{j_2(x)}{x} 
\left[-2+\frac{\ell^2}{(x_0-x)^2}\right] 
j_{\ell}(x_0-x) \nonumber\\ 
&\simeq & -\frac{1}{2} \left[{1 \over \rho_c\Om_r} 
       \ln\left(\frac{z_{\text{in}}}{z_{\text{eq}}}\right)\right] 
    {J_{\ell+3}(x_0) \over \sqrt{x_0}} \, \Pi^{\pm}(\bk) 
\label{T-E-int-soln-2} 
\end{eqnarray} 
where again $x\equiv k\eta$ and $x_0\equiv k\eta_0$, and we have evaluated 
the time integral 
using approximation (\ref{divergent-approx}). Here we have also neglected 
a term of the order of $(\ell^2/x_0^2)J_{\ell+3}(x_0)$, which in 
principle is of the same order in the above expression, but is always 
subdominant once we perform the integral over $k$. Since the power spectra 
for the $E$ polarization are parity even, 
only the parity even part of the $\Pi^\pm$ auto-correlator 
(Eq. (\ref{ft-tensor-basis})) 
contributes to the expression for $C_{\ell}^{EE}$ derivable from
Eq.~(\ref{C-power-spectrum}). Again we present here only the 
effect coming from the helical part of the magnetic field,  using 
Eq. (\ref{T-source-appendix}) 
we find ($x_0$ is renamed $x$)
\begin{eqnarray} 
C^{EE}_{(A)\ell} 
\simeq \frac{32\pi}{9}{\big[\frac{\Om_A}{\Om_r}
\ln\big({z_{\text{in}}\over 
    z_{\text{eq}}}\big)\big]^2 \over (2n_A+3)\Ga^2\left({n_A+4\over 2}\right)} 
\left({k_D\eta_0}\right)^{-3}  \int^{x_D}_0dx\, x \,J_{\ell+3}^2(x) 
\left[1+\frac{n_A-1}{n_A+4}\left(\frac{x}{x_D}\right)^{2n_A+3}\right] 
 ~. 
\label{T-E-power-spectrum-1} 
\end{eqnarray} 
The corresponding equation for $C^{EE}_{(S)\ell}$ can be found in 
Ref.~\cite{mack02}. There, a somewhat different approximation than ours 
has been used for the time integral. 
%Note that $(2\pi)^{2n_A+10}({\mathcal B}_\lambda^2 \lambda)^2 
%/\left [\Gamma^2\left(\frac{n_A+4}{2}\right) 
%k^{2n_A+6}_\lambda\right]\propto |A_0|^2$. 
 
For $n_A\geq-2$, the integral over $x$ is dominated by the upper
cutoff, $x_D=k_D\eta_0$. Using the approximation (\ref{eq:V-Bessel-integral}), we obtain 
\be
\ell^2C^{EE}_{(A)\ell} \simeq \frac{32}{9}{\big[\frac{\Om_A}
{\Om_r} 
\ln\big({z_{\text{in}}\over z_{\text{eq}}}\big)\big]^2 \over 
 (2n_A+3)\Ga^2\left({n_A+4\over 2}\right)} 
\left({\ell\over k_D \eta_0}\right)^{2} 
\times\left\{ \begin{array}{ll} 
1  & \mbox{for $n_A > -3/2$} \\ [1mm]
{n_A-1\over (n_A+4)(2n_A+4)} &  \mbox{for $-2< n_A < -3/2$}
\nonumber\\ [2mm] 
-{3\over2}
\ln \big(\frac{k_D\eta_0}{\ell^2}\big) &  \mbox{for $n_A = -2$} 
\end{array} \right. 
\label{E-power-spectrum-A} 
\ee 
The result for  $C^{EE}_{(S)\ell}$ is obtained upon replacing $n_A$ 
by $n_S$ and $\Om_A$ by $\Om_S$ (more precisely the factor 
$\Ga^2({n_A+4\over 2})$ has to be replaced by $\Ga^2({n_S+3\over 2})$
and the factor $(n_A-1)/(n_A+4)$ by $n_S/(n_S+3)$). 
For $-3<n_A<-2$, using (\ref{eq:GR-6.574.2}), we obtain 
\begin{eqnarray} 
\ell^2C^{EE}_{(A)\ell} 
&\simeq&
\frac{16\sqrt{\pi}}{9}{\big[\frac{\Om_A}{\Om_r} 
\ln\big({z_{\text{in}}\over 
  z_{\text{eq}}}\big)\big]^2\over (2n_A+3) 
\Ga^2\left({n_A+4\over 2}\right)}{\Ga(-n_A-2) \over \Ga(-n_A-{3\over
    2})}{n_A-1\over n_A+4}
\left({\ell\over k_D\eta_0}\right)^{2n_A+6}  \qquad \mbox{for $-3<n_A < -2$~.} 
\label{E-power-spectrum-B} 
\end{eqnarray} 
Again the $E$ polarization power spectrum from the 
symmetric part of the magnetic field spectrum is obtained upon 
replacement of $n_A$ by $n_S$ and $\Om_A$ by $\Om_S$. 
Similar evaluations like the ones presented in the previous paragraph 
show that 
\be 
C^{EE}_{\ell} = C^{EE}_{(S)\ell} - C^{EE}_{(A)\ell} \ge 0~. 
\ee 
 
\subsubsection{$B$ type polarization} 
 
Like for $E$ polarization, the integral solutions for $B$ polarization 
in the case of tensor perturbations are given in \cite{hu96}. In the 
approximation of instant recombination we have 
\begin{equation} 
\frac{B^{\pm}_\ell(\bk,\eta_0)}{2\ell+1}=\sqrt{2\over 
3}\int^{\eta_0}_{\eta_{\text{dec}}} 
     d\eta\,{\dot h}^{\pm }(\bk,\eta)\beta^{\pm }_{\ell}[k(\eta_0-\eta)]~, 
\label{T-B-int-soln-1} 
\end{equation} 
where 
\begin{equation} 
\beta^{\pm}_{\ell}(x)=\pm 
\frac{1}{2}\left[j'_{\ell}(x)+2\frac{j_{\ell}(x)}{x}\right] 
\simeq \pm \frac{1}{2}\left[{\ell \over x}j_\ell(x) - j_{\ell+1} (x) 
  \right] 
\quad \mbox{ for } \ell\gg 1~. 
\label{T-B-radial-fcn} 
\end{equation} 
With Eq. (\ref{T-rms-metric}) we can write the above integral in 
terms of the tensor sources $\Pi^{\pm}(\bk)$: 
\bea 
\frac{B^{\pm}_{\ell}(\bk,\eta_0)}{2\ell+1} &\simeq& 
{\pm \sqrt{6}\over \rho_c\Om_r} 
\ln\left(\frac{z_{\text{in}}}{z_{\text{eq}}}\right) 
\Pi^{\pm}(\bk)\int^{x_0}_{x_{\text{dec}}} dx\,\frac{j_2(x)}{x} 
\left[{\ell\over x_0-x}j_{\ell}(x_0-x)-j_{\ell+1}(x_0-x)\right] 
\nonumber \\ 
&\simeq& {\mp 1 \over 2\rho_c\Om_r} 
       \ln\left(\frac{z_{\text{in}}}{z_{\text{eq}}}\right)
    {J_{\ell+4}(x_0) \over \sqrt{x_0}} \Pi^{\pm}(\bk)~, 
\label{T-B-int-soln-2} 
\eea
where we have again used approximation (\ref{divergent-approx}). 
Like for the $E$ polarization, in this case also it is the parity even 
part of the magnetic source, $f(k)$, which contributes to the 
$C_\ell$. Eq. (\ref{C-power-spectrum}) takes the form 
\begin{eqnarray} 
C^{BB}_{(A)\ell} 
\simeq \frac{32\pi}{9}{\big[\frac{\Om_A}{\Om_r}
\ln\big({z_{\text{in}}\over 
    z_{\text{eq}}}\big)\big]^2 \over (2n_A+3)\Ga^2\left({n_A+4\over 2}\right)} 
\left({k_D\eta_0}\right)^{-3}
\int^{x_D}_0dx\, x \, J_{\ell+4}^2(x) 
\left[1+\frac{n_A-1}{n_A+4}\left(\frac{x}{x_D}\right)^{2n_A+3}\right] 
 ~. 
\label{T-B-power-spectrum-1} 
\end{eqnarray} 
Note that within our approximation, for $\ell\gg 1$, $C^{BB}_{(A)\ell} \simeq 
C^{EE}_{(A)\ell}$~. This is also the case for $C^{BB}_{(S)\ell}$ and 
$C^{EE}_{(S)\ell}$~, see~\cite{mack02}. Evaluating the integral using 
expressions (\ref{eq:V-Bessel-integral}) and 
(\ref{eq:GR-6.574.2}), for the different ranges of the spectral 
index $n_A$, we obtain 
\begin{eqnarray} 
\ell^2C^{BB}_{(A)\ell} &\simeq& \frac{32}{9}
{\big[\frac{\Om_A}{\Om_r} 
\ln\big({z_{\text{in}}\over z_{\text{eq}}}\big)\big]^2 \over 
 (2n_A+3)\Ga^2\left({n_A+4\over 2}\right)} 
\left({\ell\over k_D\eta_0}\right)^{2} 
\times\left\{ \begin{array}{ll} 
1 & \mbox{for $ n_A > -3/2$} \\[1mm] 
{n_A-1\over (n_A+4)(2n_A+4)} &  \mbox{for $-2< n_A < -3/2$} \nonumber
\\[2mm] 
-{3\over2} 
\ln \big(\frac{k_D\eta_0}{\ell^2}\big) &  \mbox{for $n_A = -2$} 
\end{array} \right. 
\label{B-power-spectrum-A}
\\ 
\ell^2C^{BB}_{(A)\ell} 
&\simeq& \frac{16\sqrt{\pi}}{9}{\big[\frac{\Om_A}{\Om_r} 
\ln\big({z_{\text{in}}\over 
  z_{\text{eq}}}\big)\big]^2\over (2n_A+3) 
  \Ga^2\left({n_A+4\over 2}\right)}{\Ga(-n_A-2) \over \Ga(-n_A-3/2)} 
\left({\ell\over k_D\eta_0}\right)^{2n_A+6} ~~~~~ \mbox{for}~n_A<-2~. 
\label{B-power-spectrum-B} 
\end{eqnarray} 
Again, the contributions from the symmetric part are obtained by 
replacing $\Om_A$ by $\Om_S$ and $n_A$ by $n_S$, up to factors of 
order unity and we find 
\be 
C^{BB}_{\ell} = C^{BB}_{(S)\ell} - C^{BB}_{(A)\ell} \ge 0~. 
\nonumber
\ee 
Within our approximation, which is better than a factor of $2$, we 
have $C^{BB}_{\ell} \simeq C^{EE}_{\ell}$~. From ordinary inflationary
perturbations one expects  $C^{BB}_{\ell} \simeq {8\over 13}C^{EE}_{\ell}$
for gravity waves, which is comparable to our findings.

\subsubsection{Temperature and $E$ polarization cross correlation} 
 
The symmetric part of the source term, $f(k)$, can only induce 
parity even CMB correlators. Besides the  power spectra for
temperature anisotropies and $E$ and $B$ type polarizations analyzed
in the previous subsections, it can also source the
cross-correlation between temperature anisotropy 
and $E$ polarization. 
In order to evaluate this contribution, we have to substitute into 
Eq. (\ref{C-power-spectrum}) the integral solutions for the tensor mode 
Eqs. (\ref{T-temp-int-soln-2}) and (\ref{T-E-int-soln-2}), to 
obtain: 
\begin{eqnarray} 
C^{\Theta E}_{(A)\ell}  \simeq 
\frac{32\pi}{9} 
{\big[\frac{\Om_A}{\Om_r} 
\ln\big({z_{\text{in}}\over z_{\text{eq}}}\big)\big]^2 \over 
 (2n_A+3)\Ga^2\left({n_A+4\over 2}\right)} 
\left(k_D\eta_0\right)^{-3} \, \ell^{5/2} 
\int^{x_D}_0 dx\,\frac{J^2_{\ell+3}(x)}{x^{3/2}}\left[1+\frac{n_A-1}{n_A+4} 
\left(\frac{x}{x_D}\right)^{2n_A+3}\right]~. 
\label{T-TE-power-spectrum} 
\end{eqnarray} 
We can evaluate this integral using 
(\ref{eq:GR-6.574.2}), and we find, 
\begin{equation} 
\ell^2C^{\Theta E}_{(A)\ell}\simeq \frac{16\sqrt{\pi}}{9} 
{\big[\frac{\Om_A}{\Om_r} 
\ln\big({z_{\text{in}}\over z_{\text{eq}}}\big)\big]^2 \over 
 (2n_A+3)\Ga^2\left({n_A+4\over 2}\right)} 
 {\Ga(\frac{3}{4})\over\Ga(\frac{5}{4})}
  \left({\ell\over k_D\eta_0}\right)^3 ~, \quad \mbox{ for $n_A>-3/2$}  
\label{TE-power-spectrum-A} 
\end{equation} 
and  
\begin{eqnarray} 
\ell^2C^{\Theta E}_{(A)\ell}\simeq \frac{16\sqrt{\pi}}{9} 
{\big[\frac{\Om_A}{\Om_r} 
\ln\big({z_{\text{in}}\over z_{\text{eq}}}\big)\big]^2 \over 
 (2n_A+3)\Ga^2\left({n_A+4\over 2}\right)} 
 { \Ga(-\frac{3}{4}-n_A) \over \Ga(-\frac{1}{4}-n_A) } 
 { n_A-1 \over n_A+4 }\left(\frac{\ell}{k_D\eta_0}\right)^{2n_A+6}~,
 \quad \mbox{ for $-3<n_A<-3/2$.}
\label{TE-power-spectrum-B} 
\end{eqnarray} 
 
In this case also, the contribution from the symmetric part of the 
magnetic field spectrum to the $\Theta$-$E$ correlator 
is always larger than this helical part. 
 
\section{CMB correlators caused by magnetic field helicity} 
\label{CMB-correlators} 
 
If the source (or the initial conditions) have no helical component, 
$\langle\Pi^+(\bk)\Pi^+(\bk')\rangle = \langle\Pi^-(\bk)\Pi^-(\bk')\rangle$, 
the  above correlators are the only non-vanishing ones. 
However, as soon as the tensor magnetic source spectrum has a helical
contribution (see Eq. (\ref{gt-tensor-basis})) 
$$(2\pi)^3 g(k) \equiv -{3\over 2} 
\langle \Pi^{+}{(\mathbf k}) \Pi^{+*}({\mathbf k}) - 
 \Pi^{-}({\mathbf k}) \Pi^{-*}({\mathbf k})\rangle \neq 0~,$$ 
the parity odd CMB power spectra are 
non zero. This has been observed first in~\cite{pogosian02}, where the 
vector contributions have been calculated. Here we compute the gravity 
wave contributions. 
We need again to evaluate Eq. (\ref{C-power-spectrum}). 
Taking into account that 
the gravity waves components $\dot{h}^{\pm}(\bk)$ are directly 
proportional to the source components (Eq. (\ref{T-rms-metric})), 
and considering the parity of the radial 
functions (Eqs.~(\ref{T-temp-radial-fcn}, 
\ref{T-E-radial-fcn}, \ref{T-B-radial-fcn})) 
 
\be 
j_{\ell}^{+}(x)=j_{\ell}^{-}(x), 
~~~~~~~ 
\epsilon_{\ell}^{+}(x)=\epsilon_{\ell}^{-}(x), 
~~~~~~~ 
\beta_{\ell}^{+}(x)=-\beta_{\ell}^{-}(x), 
\label{radial-fnc-parity} 
\ee 
it is clear that cross correlations between temperature and 
$B$ polarization $C_{\ell}^{\Theta B}$, and between $E$ and $B$ 
polarization $C_{\ell}^{EB}$, cannot vanish, since they are given by 
momentum integrals of $g(k)$. 
Using the expression of the tensor integral solutions $\Theta^{\pm}_{\ell}$ 
(\ref{T-temp-int-soln-2}), 
 $E^{\pm }_{\ell}$ (\ref{T-E-int-soln-2}) and 
$B^{\pm }_{\ell}$ (\ref{T-B-int-soln-2}), we can calculate 
the power spectra $C_{\ell}^{\Theta B}$ and  $C_{\ell}^{EB}$. 
 
\subsection{Temperature and $B$ polarization cross correlation} 
 
For temperature and $B$ polarization cross correlation we obtain after
integrating over time 
\be 
C^{\Theta B}_{\ell} \simeq {2\over \pi} 
\left[ {1 \over (\rho_c\Om_r)^2}
\ln^2\left(\frac{z_{\text{in}}}{z_{\text{eq}}}\right)\right]
\ell^{5/2} 
 \int_0^{k_D} dk \, k^2 \,{J_{\ell+3}(x)J_{\ell+4}(x) \over 
 x^{\frac{7}{2}}} 
\langle\Pi^{+}(k)\Pi^{+*}(k) - \Pi^{-}(k)\Pi^{-*}(k)
\rangle~. 
\ee 
The antisymmetric source function $g(k)$ is given in Eq. 
(\ref{T-A-source}), and the integral over $k$ can be calculated 
using (\ref{eq:GR-6.574.2}). Note that $g(k)$ depends 
on both the spectral indices $n_A$ and $n_S$, and we will have to 
evaluate the integral dividing the two cases 
$n_A+n_S\lessgtr-2$. We finally arrive at 
 
\begin{eqnarray} 
C^{\Theta B}_{\ell} &\simeq& - 
 \frac{2^8\pi}{9} 
{ \Om_S \Om_A \ln^2\big(\frac{z_{\text{in}}}
{z_{\text{eq}}}\big) \over 
\Om_r^2 (n_A+n_S+2) \Ga({n_A+4\over 2})\Ga({n_S+3\over 2})} 
\left(k_D\eta_0\right)^{-4} 
\,\,\ell^{5/2} \times \nonumber\\
  & & \times \int_0^{x_D} dx \, { J_{\ell+3}(x)J_{\ell+4}(x)  
  \over \sqrt{x}} \left[ 1+ \frac{n_A-1}{n_S+3}
  \left( \frac{x}{x_D} \right)^{n_A+n_S+2}\right] \nonumber \\ 
\ell^2C^{\Theta B}_{\ell} &\simeq& 
\left\{
\begin{array}{ll}
- {8\pi\sqrt{\pi/2} \Om_S \Om_A  \ln^2
\big(\frac{z_{\text{in}}} 
{z_{\text{eq}}}\big) \over \Om_r^2 (n_A+n_S+2) \Ga({n_A+4\over 2})
\Ga({n_S+3\over 2})} 
 \left(\frac{\ell}{k_D\eta_0}\right)^4  & \mbox{for $n_S+n_A>-2$} 
\label{T-TB-b} \\ [4mm] 
-{2^7\sqrt{\pi} \Om_S \Om_A \ln^2
\big(\frac{z_{\text{in}}} 
{z_{\text{eq}}}\big) \over 9 \Om_r^2 (n_A+n_S+2) \Ga({n_A+4\over 2}) 
\Ga({n_S+3\over 2})} 
\frac{\Ga\big(-\frac{n_A}{2}-\frac{n_S}{2}-\frac{1}{4}\big)} 
{\Ga\big(-\frac{n_A}{2}-\frac{n_S}{2}+\frac{1}{4}\big)} 
\frac{n_A-1}{n_S+3} 
\left( \frac{\ell}{k_D\eta_0}\right)^{n_A+n_S+6} 
& \mbox{for $-6< n_S+n_A< -2$} 
\label{T-TB-a} 
\end{array}
\right.
\end{eqnarray} 
Independently on the spectral indices, $\ell^2C_{\ell}^{\Theta B}$ 
is always negative for positive $A_0$. 

In this case of temperature and $B$ polarization cross correlation, we
have computed the spectrum (\ref{T-TB-b}) also numerically, in order
to test the reliability of our analytical estimation. The amplitude of
the numerical result is bigger than the analytic one by a factor of
two or less, so within the error we estimated for our approximations (see
Appendix B). We expect this to be one of the worst approximations due
to the relatively slow convergence of 
$\int dx J_{\ell+3}(x)J_{\ell+4}(x)/\sqrt{x}$.     
  
\subsection{$E$ and $B$ polarization cross correlation} 
 
Following the same procedure as in the previous paragraph, we can 
evaluate the $E$ and $B$ polarization cross correlation created by 
the helical part of the magnetic field. Using the formula 
(\ref{C-power-spectrum}), we get: 
\bea 
C^{EB}_{\ell} &\simeq& -{2^6\pi\over 9} 
{ \Om_S \Om_A \ln^2\big(\frac{z_{\text{in}}}{z_{\text{eq}}}\big) \over 
\Om_r^2 (n_A+n_S+2) \Ga({n_A+4\over 2})\Ga({n_S+3\over 2})} 
\left(k_D\eta_0\right)^{-4}
\times \nonumber \\
& &\times \int_0^{x_D} dx \, x^2 J_{\ell+3}(x)J_{\ell+4}(x) 
 \left[ 1+ \frac{n_A-1}{n_S+3}\left( \frac{x}{x_D} 
 \right)^{n_A+n_S+2}\right]~. 
\label{E-B-cross-corr-int}
\eea 
In the case $n_A+n_S>-2$, the integral in $x=k\eta_0$ is divergent, 
and we need to evaluate it using approximation (\ref{eq:ap2}), which
gives:
\bea \label{EB-A}
\ell^2 C^{EB}_{\ell} &\simeq & {2^5\over 9} 
{ \Om_S \Om_A  \ln^2\big(\frac{z_{\text{in}}}
{z_{\text{eq}}}\big) \over 
\Om_r^2 (n_A+n_S+2) \Ga({n_A+4\over 2})\Ga({n_S+3\over 2})} 
{(-1)^{\ell} \over k_D\eta_0}
 \sin(2x_D)\left( \frac{\ell}{k_D\eta_0}\right)^2~, \\
&& \mbox{ for $n_S+n_A>-2$.} \nonumber
\eea 
It is not possible to assign a precise value to the variable $x_D=\eta_0 
k_D$, because of the unavoidable incertitude in the estimation of the 
magnetic field damping scale, which depends on the amplitude of the 
magnetic field and is therefore smeared out over a certain range of scales. 
Therefore, we expect that the presence of the term 
$\sin(2x_D)$ most probably leads to a considerable suppression in the 
amplitude of the $E$ --- $B$ cross correlation term. 

For $n_A+n_S<-2$, the momentum integral in Eq. (\ref{E-B-cross-corr-int}) is
dominated by the second term in the square brackets, and in order to
perform the integration, we need to distinguish two different cases: 
For $-4\leq n_A+n_S<-2$, the exponent of $x$ is still
 positive, so that we have to use the approximation given in
Eq.~(\ref{eq:ap2}).  A further distinction is therefore
necessary, since the dominant term in approximation (\ref{eq:ap2})
depends on whether the exponent is above or below $1$ as discussed in
the Appendix. 
\bea
\ell^2 C_{\ell}^{EB} &\simeq& {2^5\over 9} {\Om_S \Om_A \ln^2
\big(\frac{z_{\text{in}}} 
{z_{\text{eq}}}\big) \over \Om_r^2 (n_A+n_S+2) \Ga({n_A+4\over 2}) 
\Ga({n_S+3\over 2})} 
\frac{n_A-1}{n_S+3} {(-1)^{\ell} \over k_D \eta_0} 
\sin(2x_D) \left( \frac{\ell}{k_D\eta_0}\right)^2~,
\label{EB-B}\\
&& \mbox{for $-3< n_A+n_S<-2$}; \nonumber\\
\ell^2 C_{\ell}^{EB} &\simeq& {2^5\over 9} {\Om_S \Om_A \ln^2
\big(\frac{z_{\text{in}}} 
{z_{\text{eq}}}\big) \over \Om_r^2 (n_A+n_S+2) \Ga({n_A+4\over 2}) 
\Ga({n_S+3\over 2})} 
\frac{n_A-1}{n_S+3} {(-1)^{\ell+1} \over (k_D\eta_0)^2}
 \sin(2\ell^2) 
\left( \frac{\ell^2}{k_D\eta_0}\right)^{n_A+n_S+4}~,\label{EB-C}\\
 && \mbox{for $-4< n_A+n_S<-3$} \nonumber.
\eea 
Both contributions are suppressed by the presence of the
two terms $\sin(2\ell^2)$ and $\sin(2x_D)$ since, usually one averages
over band powers in $\ell$ (for the second case) and also $x_D$ is not
a very sharp cutoff but has a certain width, as mentioned above  (for the first case).

If $-6<n_A+n_S<-4$,  the second term in the integrand of
Eq.~(\ref{E-B-cross-corr-int}) still dominates, but since the
exponent of $x$ is now negative, the integral converges
and we can make use of approximation~(\ref{eq:GR-6.574.2}).
\bea
\ell^2 C_{\ell}^{EB} &\simeq& -{2^5\sqrt{\pi}\over 9} {\Om_S \Om_A \ln^2
\big(\frac{z_{\text{in}}} 
{z_{\text{eq}}}\big) \over \Om_r^2 (n_A+n_S+2) \Ga({n_A+4\over 2}) 
\Ga({n_S+3\over 2})} 
\frac{\Ga\big(-\frac{n_A}{2}-\frac{n_S}{2}-\frac{3}{2}\big)} 
{\Ga\big(-\frac{n_A}{2}-\frac{n_S}{2}-1\big)} 
\frac{n_A-1}{n_S+3} 
\left( \frac{\ell}{k_D\eta_0}\right)^{n_A+n_S+6}~,\\
&& \mbox{for $-6<n_A+n_S<-4$.}
\eea  
This result is not suppressed by oscillations.

\section{Discussion and conclusions}
In this paper we have computed CMB anisotropies due to gravity waves
induced by a primordial magnetic field. We have mainly
concentrated on the effects of a possible helical component of the
field. Magnetic fields induce scalar, vector and tensor perturbations
which are typically of the same order. In this sense the tensor
contribution can be regarded as an order of magnitude estimate for
the full contribution.

As it has already been found in Refs.~\cite{durrer00,mack02}, the
$C_\ell$'s are proportional to
\be  \label{ampsi}
  \ell^2C_\ell \propto \left({\Om_B\over \Om_r}\right)^2
       \ln^2\left({z_{\text{in}}\over z_{\text{eq}}}\right).
\ee  
The first term is $\big({\Om_B\over \Om_r}\big)^2 \simeq
10^{-10}\left(B/10^{-8}{\rm Gauss}\right)^4$, hence for a primordial
magnetic field of the order of $B\simeq 10^{-9}$ to $10^{-8}$ Gauss we
would expect to detect its effects in the CMB anisotropy and
polarization spectrum.  Here $B =B_{k_D}=B_\la(\la k_D)^{n+3}$ is the maximum
value of the $B$-field which is always the field at the upper cutoff
scale $1/k_D$ which we also denote by $B_{k_D}$.

In Eq.~(\ref{ampsi}) $\Om_B$ stands for $\Om_S$ or $\Om_A$ and in the
above expression for $B_{k_D}$,  $n$ stands for $n_A$ or $n_S$
depending on which contribution we are considering. The second term
represents the logarithmic build up of gravity waves, 
$\ln^2\left({z_{\text{in}}/ z_{\text{eq}}}\right) \simeq 660$ to $3100$.
Here the first value corresponds to magnetic field generation at the
electroweak phase transition, $T_{\rm in} =200$ GeV and the second
value represents a possible inflationary generation at $T_{\rm in}
\simeq 10^{15}$ GeV. For scale invariant spectra, $n_A=n_S \simeq -3$,
the right hand side of Eq.~(\ref{ampsi}) gives roughly the amplitude
of the induced CMB perturbations. 

Taking into account the pre-factor $2^8\sqrt{\pi}/9$, scale
invariant magnetic fields produced at some GUT scale, 
$T\simeq 10^{15}$ GeV have to be of the order of $B \simeq {\cal B}
\simeq 10^{-10}$ Gauss to contribute a signal on the level of about 1\%
to the CMB temperature anisotropies and polarization. 
 
If the initial magnetic field is not scale invariant, the scales $k_D$
and $\eta_0$ suppress the results by factors of $1/(k_D\eta_0)$ and
$\ell/(k_D\eta_0)$ which are much smaller than unity. Note that the
reference scale $\la$ introduced in
Eqs.~(\ref{B-mean}, \ref{hel-mean}), does not enter in the final
results at all, since it is of course arbitrary.
%
% We parameterize
%the dependence on $k_D$ and $\eta_0$ in terms of the ratios  $\la/\eta_0$ and
%$k_D\la$, where $\la$ is our (arbitrary) reference scale introduced in
%Eqs.~(\ref{B-mean},\ref{hel-mean}). This is also the scale on which
%the magnetic field energy contributes a fraction $\Om_S$ or $\Om_A$
%respectively to the critical density. Hence  $\Om_S=\Om_S(\la)$ and
%$\Om_A= \Om_A(\la)$. On an arbitrary scale $L$, 
%$B_L^2 = B_\la^2(\la/L)^{n_S+3}$. Hencce $\Om_S(L) =
%\Om_S(\la)(\la/L)^{n_S+3}$. Correspondingly  $\Om_A(L) =
%\Om_A(\la)(\la/L)^{n_A+3}$.  Since $n_A \ge n_S \ge -3$, the
%contribution of $B$ to the total energy density is always dominated by
%the blue end, $k_D$. In other words     
%\bea
%\Om_S^{\rm tot} \simeq \Om_S(k_D) &=&
%\Om_S\left(k_D\la\right)^{n_S+3}\\
%\Om_A^{\rm tot} \simeq \Om_A(k_D) &=&
%\Om_A\left(k_D\la\right)^{n_A+3}~.
%\eea

As already discussed, the damping scale $k_D$ is given
by $k_D^{-1}\simeq v_A l_\ga(T_{\rm eq}) \simeq v_A \times 0.35$ Mpc, 
and $v_A$ is the
Alfv\'en velocity, $v_A^2=\langle B\rangle^2/(4\pi(\rho+p))$ for the
magnetic field averaged over a scale larger than the damping
scale. Clearly, $v_A\lsim 10^{-3}$ so that $B$ does not induce density
perturbations larger than $10^{-5}$. Therefore, the damping scale is
of the order of 1 kpc or less. The latter value is reached for maximal
magnetic fields which are of the order of $\langle B\rangle
\sim 10^{-9}$Gauss. On the other hand $a_0(\eta_0-\eta_{\rm dec})
\simeq \eta_0$ is simply the angular diameter distance to the last
scattering surface, which has been very accurately measured with the
WMAP satellite~\cite{spergel}, $\eta_0 = d_A = 13.7\pm 0.5$ Gpc.
So that $k_D\eta_0 \sim 10^7$ or even larger, depending on the
magnetic field amplitude. 
%To continue the discussion let us choose a normalization scale of
%$\la=1h^{-1}$Mpc. Using $\eta_0\simeq 6000h^{-1}$Mpc, we have 
%$\eta_0/\la \simeq 6000$. Note that $\ell_\la \simeq \pi\eta_0/\la
%\simeq 2\times 10^4$ corresponds roughly to the harmonic under which the
%scale of $1$Mpc is seen in the microwave sky. 

Our results differ somewhat, but not in a very significant way from
the results obtained in Ref.~\cite{mack02}. Since our magnetic field
spectra are either scale invariant or blue, the induced spectra
$\ell^2C_\ell$ are also either scale-invariant or blue. They grow
towards large $\ell$. It is therefore an advantage to choose $\ell$ as
large as possible. However, in our calculations we have not taken into
account the decay of gravity waves which enter the horizon before
decoupling. Our results are therfore correct only for $\ell
< \eta_0/\eta_{\rm dec} \sim 60$. 
%
%
%Silk sampling which severely reduces power from all $C_\ell$'s
%with $\ell\lsim 600$. Furthermore, we have used $\ell\la/\eta_0\ll 1$,
%which requires $\ell<<l_\la \sim 10^4$. 
To be on the safe side, we choose $\ell = 50$ in our graphics.

\begin{figure}[ht]
\begin{center}
\begin{minipage}{0.8\linewidth}
\centering
\epsfig{figure=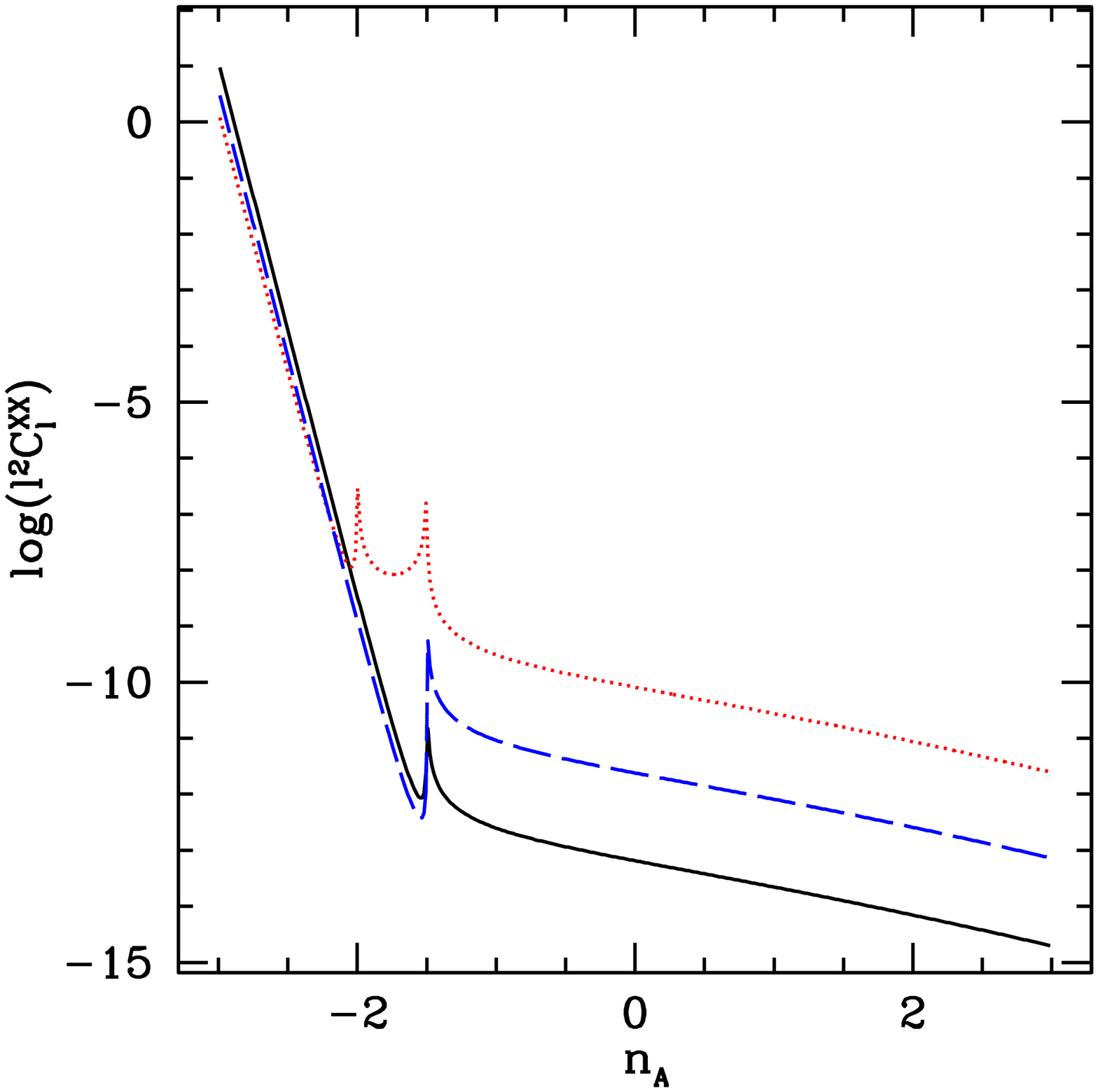,width=6.5cm} \qquad\qquad
\epsfig{figure=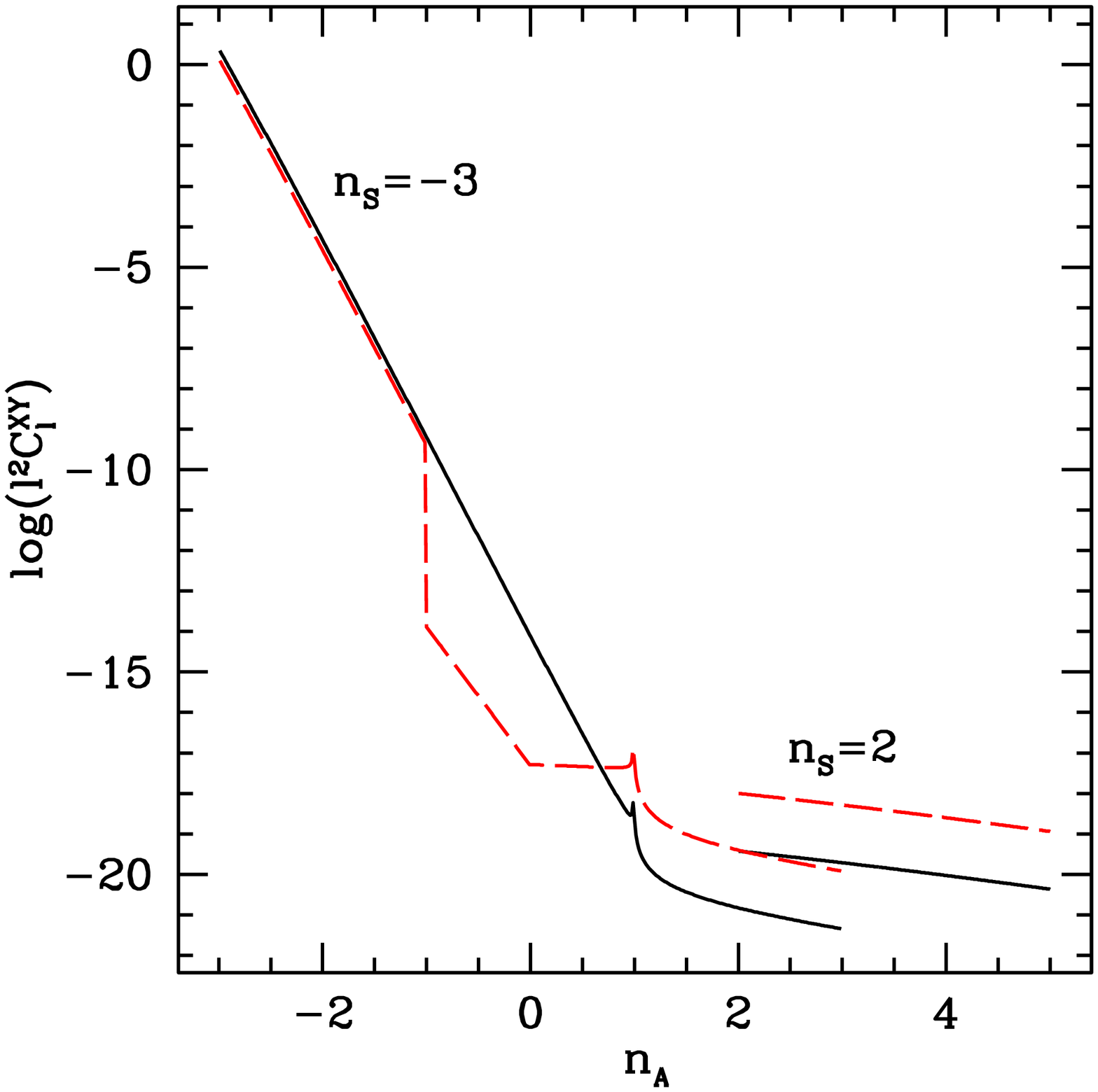,width=6.5cm}
\caption{On the top panel we show the amplitudes of the parity even
  correlators, $\ell^2C^{(\theta\theta)}_{(A)\ell}$ (solid, black), 
$\ell^2C^{(EE)}_{(A)\ell}$
  (dotted, red) and $\ell^2C^{(\theta E)}_{(A)\ell}$ (dashed, blue) as a
  function of the spectral index $n_A$ for $\ell=50$. The logarithm
  of the absolute value of $\ell^2C_{(A)\ell}^{(XY)}$ is shown in units of
$(\Om_A/\Om_r)^2\ln^2(z_{\rm in}/z_{\rm eq})$. We do not plot
  $\ell^2C^{(BB)}_{(A)\ell}$ which equals $\ell^2C^{(EE)}_{(A)\ell}$ within our
  approximation. The spikes at $n_A=-2$ for $\ell^2C^{(EE)}_\ell$ and
  at $n_A=-3/2$ are not real. They are artefacts due to the break-down
  of our approximations at these values. \\
On the bottom panel we show the corresponding parity odd correlators,
 $\ell^2C^{(\theta B)}_{(A)\ell}$ (solid, black), $\ell^2C^{(EB)}_{(A)\ell}$
  (dashed, red) in units of 
$(\Om_A\Om_S/\Om_r^2)\ln^2(z_{\rm  in}/z_{\rm eq})$ 
for $n_S=-2.99$ and $n_S=2$. In this last case, only the
allowed range $n_A\geq n_S=2$ is plotted.  
Again the spike at $n_A=1$ for $n_S=-2.99$  and the precipitous drop
at $n_A=-1$ in $\ell^2C^{(EB)}_{(A)\ell}$, are
due to the limitation of our approximation close to the transition indices.  
\label{res1} }
\end{minipage}
\end{center}
\end{figure}

In Fig.~\ref{res1}, we
show $\ell^2C_{(A)\ell}^{(XY)}$ at $\ell=50$ for the different quantities
(temperature anisotropy, $E$ and $B$ polarization and correlators)
as a function of $n_A$ with $n_S$ fixed to $2$ and $-2.99$.
We show the absolute value of the correlator in units of
$$\
\left({\Om_A \over \Om_r}\right)^2\ln^2\left({z_{\rm in} \over z_{\rm
      eq}}\right)\simeq 10^{-10}\left ({{\cal
  B}_{k_D} \over 10^{-9}{\rm Gauss}}\right)^4~,
$$
and 
$$
{\Om_A\Om_S \over \Om_r^2}\ln^2\left({z_{\rm in} \over z_{\rm eq}}\right)
 \simeq 10^{-10}\left({{\cal  B}_{k_D} \over 10^{-9}{\rm Gauss}}\right)^2
 \left({B_{k_D} \over 10^{-9}{\rm Gauss}}\right)^2~.
$$  
Note that the correlators $C^{(XX)}_A$ and  $C^{(\theta E)}_A$ are always
negative and have to be subtracted from  
$C_{(S)}^{(XY)}$ which is of the same order of magnitude or
  larger since
$\Om_S\ge\Om_A$ and $n_S\le n_A$. For the limiting case, $\Om_S\simeq\Om_A$
  and $n_S\simeq n_A$, the presence
of an helical component in the magnetic field spectrum can in principle cancel
the effect of the symmetric part on the CMB. In that very particular
case, the signature of the presence of a magnetic field will appear 
only through the parity odd correlators.

From Fig.~\ref{res1} it is clear that only for $n_{A,S}\lsim -2$ and
$\Om_A \simeq \Om_S \sim 10^{-5}$, the effect on the CMB
will be of the order of a percent or more.  In
Ref.~\cite{caprini02} it has been shown that for $n_S>-2$, magnetic
fields with $B_\la \gsim 10^{-10}$Gauss over-produce gravity waves on
small scales which is incompatible with the nucleosynthesis
bound, for $\la\sim 1$ Mpc. Here we require $ B_{k_D}\lsim
10^{-8}$ Gauss so that $\Om_B$ remains a small fraction of the
radiation density throughout. Then $B_\la = B_{k_D}(\la
k_D)^{-(n+3)}\ll  B_{k_D}$ for  $n>-2$. Therefore, by keeping  $
B_{k_D}$ sufficiently small, we automatically satisfy the bound
derived in Ref.~\cite{caprini02}.
The result is most interesting for the window of
$-3<n_S\lsim n_A\lsim -2$ and $\Om_A \simeq \Om_S
\sim 10^{-5}$, which requires ${\cal B}_{k_D} \simeq B_{k_D} \sim
10^{-10}$Gauss. Especially, if magnetic field helicity is causally
produced which implies $n_S=2$ and $n_A=3$,  this effect cannot be
observed in the CMB since
the parity violating terms are suppressed by about 15 orders of
magnitude (see lines in the lower right corner of the bottom panel of
Fig.~\ref{res1}). 

\begin{figure}[ht]
\begin{center}
\begin{minipage}{0.80\linewidth}
\centering
\epsfig{figure=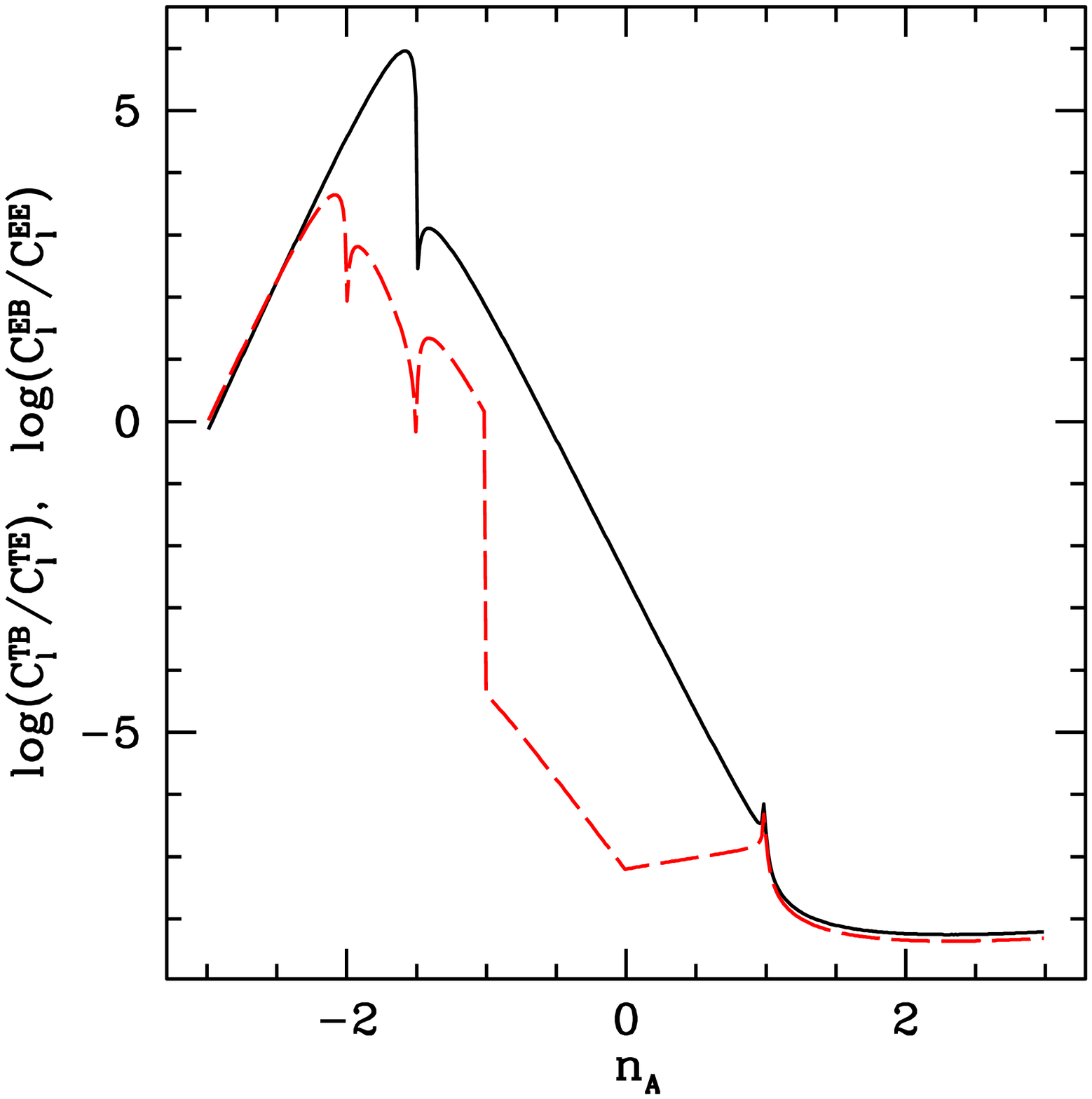,width=6.5cm}
\caption{We show the ratio of the 
  correlators, $C^{(\theta B)}_\ell/C^{(\theta E)}_\ell$ (solid, black), 
and $C^{(E B)}_\ell/C^{(EE)}_\ell$ for $n_S=-3$ as
functions of the spectral index $n_A$ for $\ell=50$. The logarithm
of the absolute value is shown in units of $\Om_A/\Om_S\le 1$. The spikes 
  visible at certain values of the spectral index $n_A$ are mainly due
  to our relatively crude approximations.
\label{res2} }
\end{minipage}
\end{center}
\end{figure}

In Fig.~\ref{res2} we show the ratio $C_{(A)\ell}^{\theta
B}/C_{(A)\ell}^{\theta E}$
for $n_S=-3$ as function of $n_A$. Again, we are mainly interested in
the part of the graph with $-3<n_A<-2$, where this ratio raises from
the  order unity to about $10^5$. Hence if a close to maximal helical
magnetic field, with a spectrum not too far from scale invariant, 
$-3<n_S<n_A<-2$ is produced in the early universe, it is more
promising to search for its parity violating terms than for the parity
even contributions.

We can conclude that helical magnetic fields with a spectrum close to
the scale invariant value, $-3 < n_S\simeq n_A\lsim -2$ and close to
maximal amplitudes on small scales, $B_{k_D}\gsim 10^{-10}$ Gauss can lead to observable
parity violating terms $C^{\theta B}$ and $C^{EB}$ in the CMB.
Such magnetic fields might in principle be produced during some
inflationary epoch where the photon is not minimally coupled or via
its coupling to the dilaton (see~\cite{inflat,stringcos} for various
proposal of magnetic field production during an inflationary phase). 
However, so far no concrete proposal has led to $n_{S,A} \simeq -3$,
nor to the creation of a helical term.
As we have shown, the effect is largely suppressed and clearly unobservable for
causally produced magnetic fields, \eg, during the electroweak phase
transition or even later.

Nevertheless, our calculation also demonstrates the effect of
parity violating processes during inflation which may lead to a non-vanishing
helical component of gravity waves, $\HH \neq 0$, see
Eq.~(\ref{gw-cross-basis}). In this case the above calculation can be trivially
repeated and will result in non-vanishing parity violating CMB correlators,
$C^{\theta B}\neq 0$ and $C^{EB}\neq 0$. We think that already this remark,
together with our knowledge that at least at low energies, nature does
violate parity, should be sufficient motivation to derive experimental
limits on these correlators. 

\acknowledgments 
We have benefited from discussions with Pedro Ferreira, Grigol
Gogoberidze, Arthur Kosowsky, Andy Mack, Bharat Ratra and Tanmay Vachaspati. 
We thank Antony Lewis for signaling the normalisation error in
Eq.~(\ref{source-tensor}). 
C.C. is grateful to Guillaume van Baalen and Thierry 
Baertschiger for assistance
with the numerical codes. C.C. and T.K. thank Geneva University for
hospitality. We acknowledge  
financial  support form the TMR network CMBNET and from the Swiss 
National Science Foundation.

\newpage

\appendix 
 
\section{The source for gravity waves } 
\label{sec:tau} 
 
In this appendix we present some details on how to compute 
the gravity waves source functions  $f(k)$ and $g(k)$. 
The first step is to evaluate the two point correlator of the 
magnetic field stress-energy tensor (\ref{tau-2corr}): using Wick's theorem 
(\ref{Wick}) and definition (\ref{spectrum}), after a 
longish but simple calculation we obtain 
\bea \langle 
\tau_{ij}({\mathbf k})\tau^{*}_{lm}({\mathbf k^\prime}) 
\rangle &=&\frac{1}{4} {1 \over (4 \pi)^2} \delta(\bk - 
\bk^\prime) \int d^3p \left(\right. S(p)S(|\bk-\bp|) 
[(\delta_{il}-\hat{p}_i\hat{p}_l)(\delta_{jm} -(\widehat{ 
\bk-\bp})_j(\widehat{\bk-\bp})_m) \nonumber 
\\ 
& & 
+(\delta_{im}-\hat{p}_i\hat{p}_m)(\delta_{jl} 
-(\widehat{\bk-\bp})_j(\widehat{\bk-\bp})_l)] 
\nonumber 
\\ 
& & 
- A(p)A(|\bk-\bp|) 
[\epsilon_{ilt}\epsilon_{jmr}{\hat p}_{t} 
(\widehat{\bk-\bp})_{r} 
+\epsilon_{imf}\epsilon_{jlg} {\hat p}_{f} 
(\widehat{\bk-\bp})_{g}] 
\nonumber 
\\ 
& & 
+ i S(p)A(|\bk-\bp|) 
[\epsilon_{jmr}(\delta_{il} - {\hat p}_i {\hat p}_l) 
(\widehat{\bk-\bp})_{r} 
+\epsilon_{jlg}(\delta_{im} - {\hat p}_i {\hat p}_m) 
(\widehat{\bk-\bp})_{g}] 
\nonumber 
\\ 
& & 
+ i A(p)S(|\bk-\bp|) 
[\epsilon_{ilt} (\delta_{jm} - {\widehat{\bf (k-p)}}_j 
{\widehat{\bf (k-p)}}_m) {\hat p}_{t} 
+ \epsilon_{imf} (\delta_{jl} - {\widehat{\bf (k-p)}}_j 
{\widehat{\bf (k-p)}}_l) {\hat p}_{f} ] 
\left. \right\} \nonumber\\ 
&& + \cdots\de_{ij} +\cdots\de_{lm}  ~. \label{tau-appendix} 
\eea 
The  isotropic tensor spectrum in the 
case of a magnetic field spectrum without helicity term 
is derived in~\cite{durrer00}. 
Here we concentrate on the source terms 
which contain the helical part of the magnetic 
field spectrum. 
 
By acting with tensor projector on (\ref{tau-appendix}), we find 
expressions (\ref{tensor-source-2}) and (\ref{tensor-source-3}) 
for the symmetric and helical parts of the source spectrum. 
Taking into account that the angle 
$\beta={\hat \bk}\cdot(\widehat{{\bf k-p}})={{k - p\gamma} 
\over {\sqrt{k^2-2kp\gamma +p^2}}}$, we can rewrite the two expressions 
which contain $A(k)$ in the form 
 
\bea 
f^{A}(k) &=& {8 \over (4\pi)^5} \int d^3p 
A(p)A(|\bk-\bp|) {{\gamma \cdot (k-p\gamma)} \over 
{\sqrt{k^2-2kp\gamma +p^2}}} \label{tensor-appendix} \\ 
g(k) &= & {4 \over (4\pi)^5} \int d^3p \left[ 
S(p)A(|\bk-\bp|) {(k-p\gamma) (1+\gamma^2) \over 
\sqrt{k^2-2kp\gamma +p^2}} +A(p)S(|\bk-\bp|) \left( 
2\gamma - {\gamma p^2 (1-\gamma^2) \over k^2-2kp\gamma +p^2} 
\right) \right] \label{tensor-appendix1}~. 
\eea 
The contribution to $f(k)$ from $S$ alone is computed in 
Ref.~~\cite{durrer00}. There one finds 
\be 
f^{S}(k) = {2 \over (4\pi)^5} \int d^3p 
S(p)S(|\bk-\bp|)(1+\gamma^2)\left(1+{(k-p\gamma)^2  \over 
k^2-2kp\gamma +p^2}\right)~. 
\ee 
We can now substitute the power law Ansatz~ (\ref{Sspec},\ref{Aspec}) 
for $S$ and $A$ in  these expressions and try to calculate the integrals. 
The integration over $\gamma={\hat \bk}\cdot {\hat \bp}$ is elementary, 
using 
\bea 
\int d\gamma\,(k^2+p^2-2kp\gamma)^{\al \over 2} 
&=&-\frac{1}{kp(\al+2)}(k^2+p^2-2kp\gamma)^{\al+2 \over 2} 
\nonumber \\ 
\int d\gamma\,\gamma^m\,(k^2+p^2-2kp\gamma)^{\al \over 2} 
&=& - \frac{\gamma^m}{kp(\al+2)}(k^2+p^2-2kp\gamma)^{\al+2 \over 2} 
+ \frac{m}{kp(\al+2)} \int d\gamma\,\gamma^{m-1}\, 
(k^2+p^2-2kp\gamma)^{\al+2 \over 2} 
\label{gamma-integral}~. 
\eea 
This last integration by parts has to be performed in the worst cases 
three times, reducing the power $m$ of $\gamma$ from $3$ down to $0$. 
 
Since we are integrating $\gamma$ over the interval $[-1,1]$, we get a series 
of $m+1$ terms of the form 
\be 
{(k+p)^{\al + 2n}\pm|k-p|^{\al+2n} \over (k\,p)^{n}}~, 
\label{gamma} 
\ee 
with $n=1,2,... (m+1)$. 
To evaluate the integral over $p$, we can expand those terms 
using the binomial decomposition 
$(1+x)^{\al}= 1 +\al x + \al(\al-1)x^2 + \cdots $. 
Since, in general, the value of the exponent $\al$ is not an integer, we 
need to truncate the series somewhere, which is well justified only 
if $x\ll1$. 
To achieve this, we split  the integral into two contributions, 
$\int_0^{k_D} = \int_0^{k} 
+ \int_{k}^{k_D}$. In the first term $p/k<1$, while 
 in the second  $k/p<1$, which allows us to approximate 
Eq. (\ref{gamma}) truncating the binomial series at the second term, 
\begin{equation} 
(k+p)^{\al}-|k-p|^{\al}\simeq\cases{ 
2 \al k^{\al-1} p + {1 \over 3} \al(\al-1)(\al-2) k^{\al-3} p^3 
& $p<k$\cr 
2 \al p^{\al-1} k + {1 \over 3} \al(\al-1)(\al-2) p^{\al-3} k^3 
& $p>k$ \cr} 
\label{eq:k-p-relation} 
\end{equation} 
and 
\begin{equation} 
(k+p)^{\al}+|k-p|^\al \simeq\cases{2k^\al +  \al(\al-1) 
    k^{\al -2} p^2 
& $p<k$\cr 
2p^\al +  \al(\al-1) p^{\al-2} k^2 
& $p>k$ ~. \cr} 
\label{eq:k+p-relation} 
\end{equation} 
We then perform the integration over $p$. For each contribution we 
keep only the terms which, depending on the value of the spectral 
index, may dominate the result. So, we finally obtain, for $k<k_D$ 
%After some long calculations, 
\bea 
f(k) &\simeq& 
\frac{\la^3}{2(2n_S+3)} 
\left[\frac{B^2_\lambda} 
{2\Gamma\left(\frac{n_S+3}{2}\right)}\right]^2 
\left( (\la k)^{2n_S+3}_D+\frac{n_S}{n_S+3}(\la k)^{2n_S+3}\right) - 
\nonumber 
\\ 
& & \mbox{} 
-\frac{2\la^3}{3(2n_A+3)} \left[\frac{{\cal B}^2_\la}
{2\Gamma\left(\frac{n_A+4}{2}\right)}\right]^2 
\left((\la k)^{2n_A+3}_D+\frac{n_A-1}{n_A+4}(\la k)^{2n_A+3}\right)
\label{T-source-appendix} \\ 
&\simeq& {\mathcal A}_S \la^{2n_S+3}\left(k^{2n_S+3}_D+\frac{n_S}{n_S+3} 
\,k^{2n_S+3}\right) 
- {\cal A}_A\la^{2n_A+3} \left(k^{2n_A+3}_D+\frac{n_A-1}{n_A+4} \, 
k^{2n_A+3}\right)\\ 
g(k) & \simeq & 
\frac{4}{3}\frac{\la^{4} k}{(n_S+n_A+2)} \left[ 
\frac{B^2_\lambda} 
{2\Gamma\left(\frac{n_S+3}{2}\right)}\right] 
\left[\frac{ {\mathcal B}^2_\lambda} 
{2\Gamma\left(\frac{n_A+4}{2}\right)}\right] 
\left( (\la k_D)^{n_S+n_A+2} + \frac{n_A-1}{n_S+3}(\la k)^{n_S+n_A+2} 
\right)
\label{T-source-hel-appendix}\\ 
&\simeq&  {\cal C}k\la \,(\la k_D)^{n_S+n_A+2}\, \left[1 + \frac{n_A-1}{n_S+3} 
\left(\frac{k}{k_D}\right)^{n_S+n_A+2}  \right]~, 
\eea 
where the coefficients are given by the magnetic field amplitudes at 
scale $\la$: 
\bea 
{\cal A}_S &\simeq& \frac{\la^3}{2(2n_S+3)} 
\left[\frac{B^2_\lambda} 
{2\Gamma\left(\frac{n_S+3}{2}\right)}\right]^2 \label{T-source-new}\\ 
{\cal A}_A &\simeq& \frac{2\la^3}{3(2n_A+3)} \left[\frac{
{\mathcal B}^2_\lambda} 
{2\Gamma\left(\frac{n_A+4}{2}\right)}\right]^2 \\ 
{\cal C} &\simeq&  \frac{4}{3}\frac{\la^3}{(n_S+n_A+2)} \left[ 
\frac{B^2_\lambda} 
{2\Gamma\left(\frac{n_S+3}{2}\right)}\right] 
\left[\frac{{\mathcal B}^2_\lambda} 
{2\Gamma\left(\frac{n_A+4}{2}\right)}\right] ~. \label{T-source-hel-new} 
\eea 
The first part of $f(k)$, which is the contribution from the 
symmetric part of the magnetic field power spectrum, has been taken 
from \cite{durrer00,mack02}. The singularities  at $n_S,~n_A = 
-3/2$ respectively and at $n_S+n_A=-2$ are removable. 
 
\section{useful mathematical relations} 
\subsection{Integrals of Bessel functions} 
 
In Section~\ref{CMB-fluctuations}, we use approximate solutions 
for the three integrals 
\be 
\int^{x_0}_{x_{\text{dec}}} 
dx\,\frac{j_2(x)}{x}\frac{j_{\ell}(x_0-x)}{(x_0-x)^2}~,~~~~~~~~~~ 
\int^{x_0}_{x_{\text{dec}}} dx\,\frac{j_2(x)}{x} \frac{j_{\ell}(x_0-x)}
{(x_0-x)}~,~~~~~~~~~~ 
\int^{x_0}_{x_{\text{dec}}} dx\,\frac{j_2(x)}{x} j_{\ell}(x_0-x)~. 
\label{three-int-appendix} 
\ee 
These integrals are solvable only by numerical method. 
However, the aim of this paper is to give an approximate analytic
result. In this appendix we  therefore derive and test analytic
approximations to the above integrals. To achieve this, 
we first modify them slightly, in order to make them solvable analytically.
Then, we adjust the result obtained in this way by comparing 
it with the exact numerical integration. 
 
Let us concentrate, as an example, 
on the first integral. We first perform a variable transform to
$y=x_0-x$. The integration boundaries then become $0$ and
$x_0-x_{\text{dec}}$. Below, we derive an approximation for
\[
\int_0^{x_0}\frac{j_2(x_0-y)}{x_0-y} \frac{j_{\ell}(y)}{y^n}dy
\]
Since Bessel functions change on a scale $\De y \sim 1$, this
approximation is good  for the integrals in 
Eq.~(\ref{three-int-appendix}) if $x_{\text{dec}} < 1$.  
After the integration over x in Eq.(\ref{three-int-appendix}) we have
to perform an integration over $k$. For $\ell$ fixed, this integral is
either dominated by the contribution art $k\eta_0=x_0=\ell$ or at the
upper cutoff, $k_D$. For the
integrals which are dominated at $x_0=k\eta_0 \sim \ell$, the
inequality $x_{\text{dec}} < 1$ is
equivalent to $\ell \simeq  x_0 \simeq 60x_{\text{dec}} \lsim 60$.
In some cases, however, our integral over $k$ is dominated at the upper
cutoff $k_D$ with $\eta_{\text{dec}}k_D\gg 60$ and of course also
$\eta_0k_D\gg 60$. Since for $\ell\simeq 60$, the dominant
contribution to the integral comes from $y\lsim 60$, our inaccuracy
of the boundary will not invalidate the approximation also for this case.  

The approximation in the upper boundary of the integral,
$x_0 - x_{\text{dec}} \simeq x_0$ makes us miss the characteristic
decay of fluctuations on angular scales corresponding to $\ell\gsim 60$.

%Then we use that
%$x_0/x_{\text{dec}}=\eta_0/\eta_{\text{dec}}\simeq 60$. 
%Since the integrated function takes its maximum value around 
%$y=x_0-x\simeq\ell$, when taking as interval of integration $[0,x_0]$, we 
%can overestimate the result only if $x_0-x_{\text{dec}}<\ell<x_0$. This means 
%that for modes which are super horizon at decoupling, 
%$k\eta_{\text{dec}}=x\lsim 1$, neglecting $x_{\text{dec}}$ is justified,
%however, nothing can be said for the modes which are inside the
%horizon at decoupling, 
%$\eta_{\text{dec}}^{-1}<k<k_D$, where $k_D$ is the upper cutoff. 
%In deriving our approximations, we have always neglected
%$x_{\text{dec}}$ in the lower bound of 
%the integral. This means then, that our approximations are only reasonable
%for modes outside the horizon at decoupling. We therefore miss the
%typical decay of the tensor spectrum for $\ell>60$, and evaluate the
%contribution from magnetic field only in the part of the spectrum
%of the integrated Sachs-Wolfe plateau.    

To make the first integral in Eq. (\ref{three-int-appendix}) 
solvable analytically, we now modify the 
powers of $y$ and $x_0-y$. Taking into account that the 
spherical Bessel Function $j_\nu(x)$ has its maximum value at 
$x\simeq\nu$, we make the attempt: 
\bea 
\int^{x_0}_0 
dx\,\frac{j_2(x)}{x}\frac{j_{\ell}(x_0-x)}{(x_0-x)^2}&=& 
\frac{\pi}{2} \int^{x_0}_0 dx\,\frac{J_{5/2}(x)}{x^{3/ 
2}}\frac{J_{\ell+{1 \over 2}}(x_0-x)}{(x_0-x)^{5/2}} \nonumber \\ 
&\simeq& \frac{\pi}{2} \sqrt{\frac{2}{5\,\ell}} \int^{x_0}_{0} 
dx\,\frac{J_{5/2}(x_0-y)}{x_0-y} 
\frac{J_{\ell+{1 \over 2}}(y)}{y^{2}}~. \\
& \simeq & {\pi\over 5} \sqrt{\frac{2}{5\,\ell}}{J_{\ell+3}(x_0)\over x_0^2} ~.
\eea 
For the last equality, we have used 6.581.2 of \cite{gradshteyn94}~, 
\bea 
\int^a_0dx\,x^{b-1}(a-x)^{-1}J_p(x)J_q(a-x) 
&=&\frac{2^b}{aq}\sum^\infty_{m=0} 
\frac{(-1)^m\,\Gamma(b+p+m)\,\Gamma(b+m)}{m!\,\Gamma(b)\,\Gamma(p+m+1)} 
\,(b+p+q+2m)\,J_{b+p+q+2m}(a)~, 
\nonumber 
\\ 
& &\mbox{}(\text{Re}\,(b+p)>0,\,\text{Re}\,q>0) 
\label{eq:GR-6.581.2} 
\eea 
and the recurrence relation 
$J_{\nu-1}(x)+J_{\nu+1}(x)=\frac{2\nu}{x}J_{\nu}$ (9.1.27 of 
\cite{abramowitz72}),  keeping only the highest order terms in $\ell$.
We can now compare this approximated analytic result with an exact 
numerical integration. Since the analytic result is again a Bessel 
function divided by a power law, it has a maximum at $x_0\simeq\ell$, and 
its envelope has a power law decay for $x_0>\ell$. This two 
characteristics are very well reproduced by the numerical result, 
which however decays somewhat faster; it turns out that a better 
approximation is 
\be 
\int^{x_0}_0 dx\,\frac{j_2(x)}{x}
\frac{j_{\ell}(x_0-x)}{(x_0-x)^2} 
\simeq \frac{1}{3} \sqrt{\frac{3\,\ell}{2}}\, \frac{J_{\ell+3}(x_0)} 
{x_0^3}~. 
\label{approx1} 
\ee 
To estimate the goodness of our approximation, let us now take into 
account the integration over
 $k$, as  in Eq. (\ref{C-power-spectrum}). What we are 
finally interested in is (Eq. (\ref{T-temp-A-power-spectrum-1})) 
\be 
\int_0^{x_D}dx_0 \,x_0^2 \, {J^2_{\ell+3}(x_0)\over x_0^6}
\left[1+\frac{n_A-1}{n_A+4} \, 
\left({x_0\over x_D}\right)^{2n_A+3}\right]~. 
\label{int-appendix} 
\ee 
As already discussed in the main text, this integral is always convergent 
and dominated by the contribution around $x_0\simeq\ell$~: we should
therefore make sure that our approximation is good around that
value. We have that for $\ell=30$, our approximation underestimates
the numerical result by about factor of two; 
for $\ell=40$, the error reduces to  $15\%$, and is always smaller 
for larger values of $\ell$. 

\begin{figure} 
 \begin{minipage}{0.85\linewidth}
\centering
\epsfxsize=6cm 
\epsfig{file=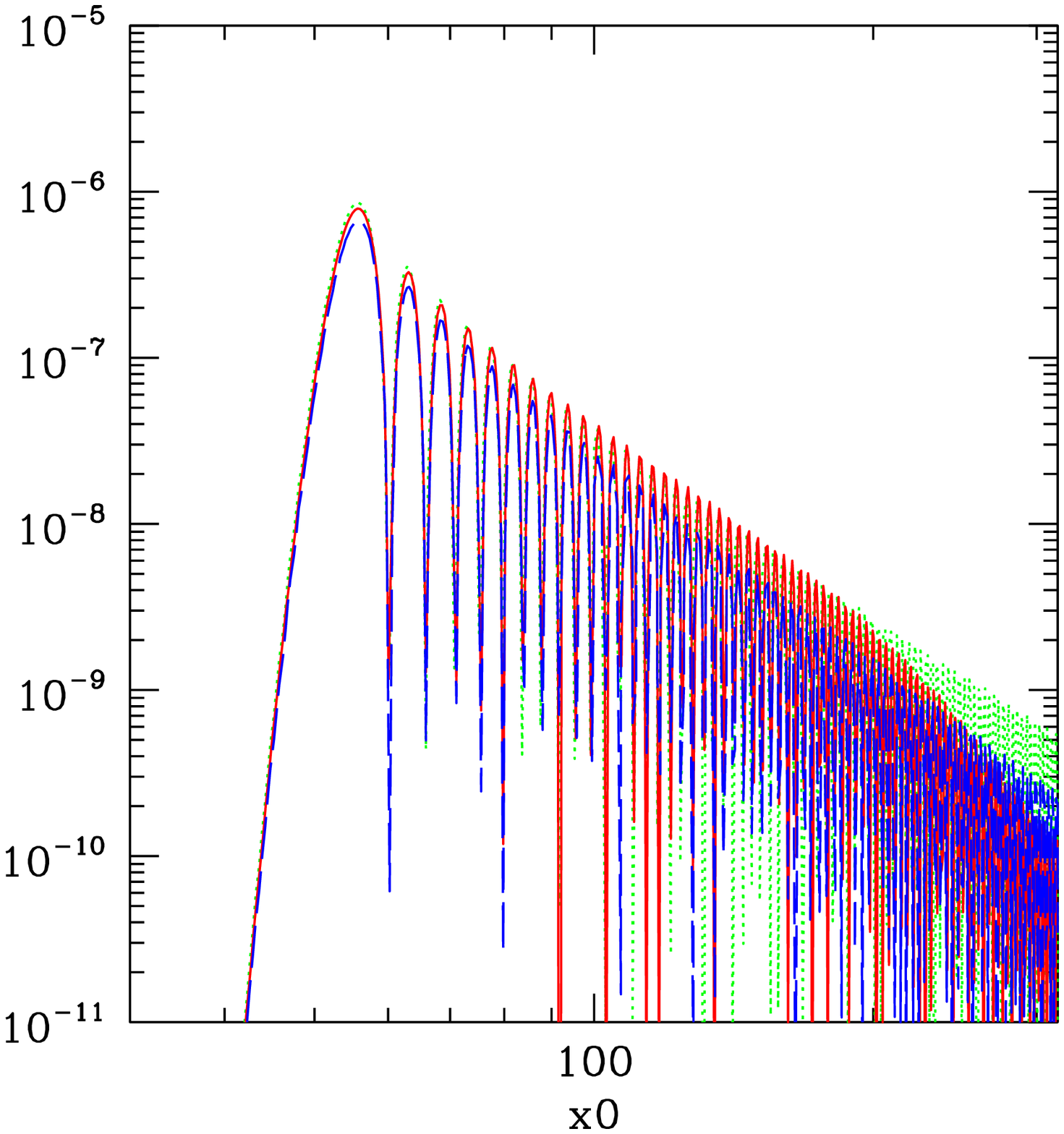,width=7.5cm} 
\epsfxsize=6cm 
\epsfig{file=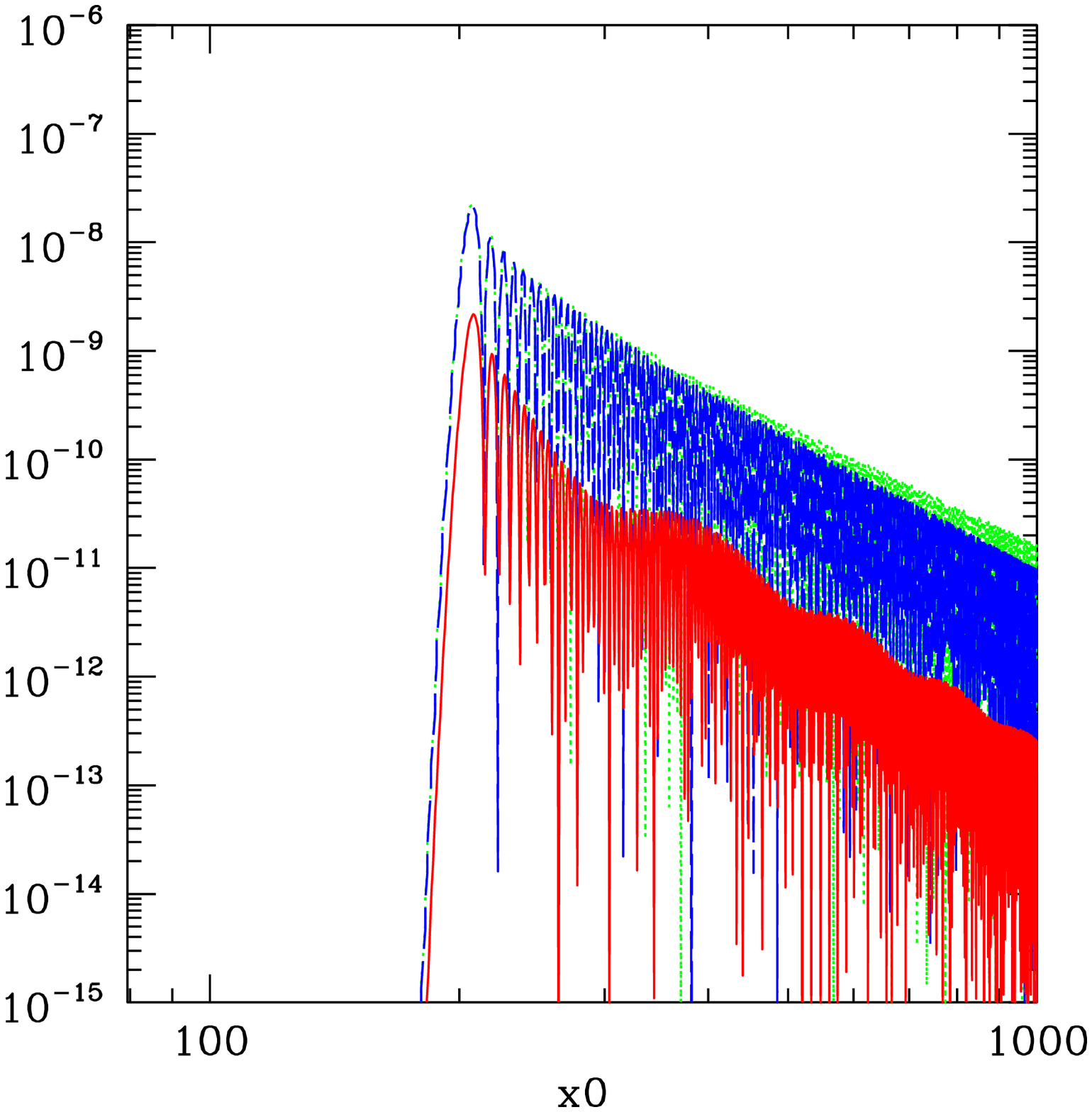,width=7.5cm} 
\caption{In both panels, as a function of $x_0$: the green dotted line
  shows the numerical  value of the integral in (\ref{approx1}), the
  blue, long dashed line shows  the analytic approximation (right hand
  side of Eq. (\ref{approx1})), and the red, solid 
  line shows the numerical value of integral (\ref{approx1}) if
  $x_{\rm dec}$ is not put to zero. All these functions are squared,
  and multiplied by $x_0^3$: this gives us an indication of the result, after 
  the integration over $x_0$, as stated in 
  Eq. (\ref{int-appendix}). In the left panel $\ell=50$, in the right
  panel $\ell=200$. First of all, we note that it appears
  clearly that the value of the 
  integrals is dominated at $x_0\simeq\ell$, and that the function goes 
  to zero quicker than $x_0^{-3}$, which justifies our 
  approximation $x_D\rightarrow\infty$ and the use of formula 
  \ref{eq:GR-6.574.2}. Secondly, we note that for $\ell=50$ and
  $x_0\sim \ell$, our approximation (blue, long-dashed) is good for both the
  integrals. However, if $\ell=200$, the approximation overestimate
  the correct numerical result by about a factor of ten.} 
\label{fig1} 
\end{minipage}
\end{figure}

Fig.~\ref{fig1} shows the numerical result for the integral in
(\ref{approx1}) (green, dotted line), together with its analytical
approximation (the right hand side of Eq. (\ref{approx1}), blue and
long dashed) and a numerical evaluation of the same integral when
$x_{\rm dec}$ is not set to zero (red, solid). For small values of
$\ell$ (in the left hand panel of Fig.~\ref{fig1}, $\ell=50$), 
Eq.~(\ref{approx1}) is a good approximation in the region
$x_0\simeq\ell$. However, if $\ell>60$ setting
$x_{\rm dec}\rightarrow 0$ causes a large overestimation of the result. In
the right hand panel of Fig.~\ref{fig1} it is shown that, for $\ell=100$, the
difference between the integral with lower bound $0$ and the one with
lower bound $x_{\rm dec}$ is of more than a factor of ten. Consequently,
as already stated before, we can rely on all our approximations only
for $\ell\lsim 60$.

We proceed now to evaluate integral (\ref{int-appendix}). Since
$x_D=k_D\eta_0\gsim 10^6$, for  $\ell\lsim 60$, 
integral~(\ref{int-appendix}) can be calculated in the limit $x_D
\rightarrow \infty$, using formula 6.574.2 of \cite{gradshteyn94}:  
\bea 
\int^{\infty}_0dx\,J_p(x)J_q(x)x^{-b}&=& 
\frac{\Gamma(b)\Gamma\left(\frac{p+q-b+1}{2}\right)} 
{2^b\Gamma\left(\frac{-p+q+b+1}{2}\right) 
\Gamma\left(\frac{p+q+b+1}{2}\right) 
\Gamma\left(\frac{p-q+b+1}{2}\right)} \label{eq:GR-6.574.2} \\ 
&&\mbox{}(\text{Re}\,(p+q+1)>\text{Re}\,b>0)~. \nonumber
\eea 
This approximation is used for example in 
Eqs. (\ref{T-temp-power-spectrum-A}, \ref{T-temp-power-spectrum-B}). 
 
With the same procedure we can approximate the second integral of 
Eq. (\ref{three-int-appendix}), for which we find ($\ell\lsim 60$)
\bea 
\int^{x_0}_{x_{\rm dec}} dx\,\frac{j_2(x)}{x} 
\frac{j_{\ell}(x_0-x)}{(x_0-x)} 
\simeq \frac{1}{3} \sqrt{\frac{3\,\ell}{2}} 
\frac{J_{\ell+3}(x_0)}{x_0^2}~. 
\label{approx2} 
\eea 
This approximation underestimates the numerical result with an error 
of about $40\%$ for $\ell=30$, which reduces to $20\%$ at $\ell=60$. In this 
case also, the integral over $x_0$ is convergent, and we can 
proceed as before.

\begin{figure}
 \begin{minipage}{0.85\linewidth}
\centering
\epsfig{file=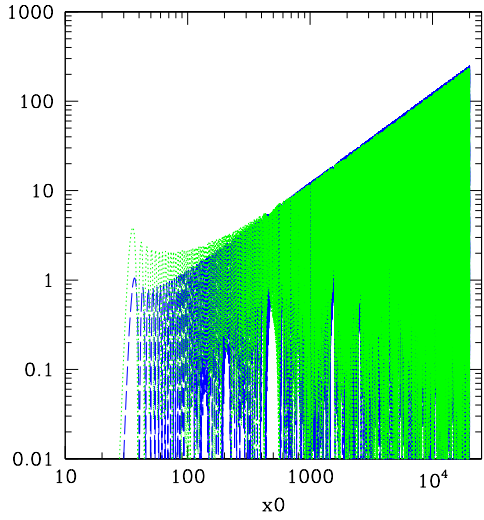,width=7.5cm} 
\caption{We plot the value of integral 
  (\ref{divergent-approx})  squared and multiplied by 
  $x_0^3$ as function of $x_0$, for $\ell=30$.
  The green, dotted line represents again the numerical
  result ($x_{\rm dec}\rightarrow 0$), and the blue, long dashed 
  line is the analytic approximation.
  In this case the slope is positive, and hence the integral 
   $dx_0/x_0$ of this function is dominated by the upper cutoff.} 
  \label{fig2} 
\end{minipage}
\end{figure}

The situation is different for the third integral of 
Eq. (\ref{three-int-appendix}). In this case, the numerical result is 
approximated by the following function ($\ell\lsim 60$): 
\be 
\int^{x_0}_{x_{\rm dec}} dx\,\frac{j_2(x)}{x} j_{\ell}(x_0-x) 
\simeq \frac{1}{3} \sqrt{\frac{2}{5}} \, \frac{J_{\ell+3}(x_0)} 
{\sqrt{x_0}}~. 
\label{divergent-approx} 
\ee 
It is clear that if we insert this function in an integral like 
(\ref{int-appendix}) we cannot perform the limit $x_D\ra \infty$ since
this integral is dominated at the upper cutoff. Consequently, we
need a good approximation for the behavior of the integral for
 large values of $x_0 \ra x_D$. In this 
case, we no longer require our approximation to be accurate at 
$x_0\simeq\ell$, but we concentrate on its behavior for high values of 
$x_0$, which will dominate in the integral over $x_0$. 
Fig.~\ref{fig2} shows the approximation for $\ell=30$, which
overestimate the numerical result by an error within $1\%$. 
 
We also have to evaluate the integral over $x_0^2\,dx_0$ 
of the square of (\ref{divergent-approx}), which we encounter in two
different cases. The first (see
Section~\ref{sub-CMBpolarization}) is of the kind  
$\int^{x_D}_0dx\,x^pJ^2_{\ell}(x)$. For $p<0$ this integral converges
and we may evaluate it in the limit $x_D\ra \infty$, in which it is of
the form (\ref{eq:GR-6.581.2}).  For $p>0$ and $x_D\gg
\ell^2$, the integral can be approximated using the  
asymptotic expansion of $J_\ell(x)$ for large arguments~\cite{abramowitz72}, 
$J_{\ell}(x)\sim\sqrt{2/(\pi x)}\cos[x-(2\ell+1)\pi/4]$. Approximating the 
oscillations by a factor of $1/2$, we obtain 
\be 
\int^{x_D}_0dx\,x^pJ^2_{\ell}(x)\simeq\int^{x_D}_{\ell^2}dx\,x^pJ^2_{\ell}(x) 
\simeq 
\cases{\frac{x^p_D}{p\pi},                     & $p>0$\cr 
       \frac{1}{\pi}\ln\left(\frac{x_D}{\ell^2}\right), & $p=0$~.\cr} 
\label{eq:V-Bessel-integral} 
\ee 
 
For the second case, $\int_0^{x_D} dx\, x^pJ_\ell(x)J_{\ell+1}(x)$, 
which we encounter in 
Section~\ref{CMB-correlators}, we use again the large argument
approximation for the Bessel functions,  for $x\gg\ell^2$, 
\bea 
J_\ell(x)J_{\ell+1}(x) &\simeq& {2\over \pi 
  x}\cos\left(x-(2\ell+1){\pi\over
  4}\right)\cos\left(x-(2\ell+3){\pi\over 4}
\right) 
  = {2\over \pi 
  x}\cos\left(x-(2\ell+1){\pi\over
    4}\right)\sin\left(x-(2\ell+1){\pi\over 4}
\right) \nonumber \\ 
&=&{1\over \pi x}\sin \left(2x-\left(\ell+{1\over 2}\right)\pi\right) 
= {(-1)^{\ell+1} \over \pi x}\cos(2x)~, 
\eea 
so that for $p>0$
\be \int_0^{x_D}dx\,x^pJ_{\ell}(x)J_{\ell+1}(x)\simeq 
{(-1)^{\ell+1}\over 
  \pi}\int^{x_D}_{\ell^2}dx\,x^{p-1}\cos\left(2x\right) 
  \simeq {(-1)^{\ell+1} \over 2\pi}\left(x_D^{p-1}\sin(2x_D)-
\ell^{2p-2}\sin(2\ell^2)\right) ~. \label{eq:ap2} 
\ee 
In the limits to which we have restricted ourselves, we always have  
$x_D\gg\ell^2$. Consequently, 
the dominant contribution in the last expression can be given either by
the first term in the bracket, if $p>1$, or by the second term, if $p<1$.
Numerical checks show that the approximation is good for $p>1$, but it
is  rather poor in the second case, $p<1$. Since we
shall not be very much interested in this case, we do not go any
further in this work. 

When evaluating expression (\ref{eq:GR-6.574.2}), we  often also use 
\bea 
\Gamma(2x)&=&\frac{2^{2x-1}}{\sqrt{\pi}}\Gamma(x) 
\Gamma\left(x+\frac{1}{2}\right) \nonumber \\
\frac{\Ga(x+a)}{\Ga(x+b)}&\sim & x^{a-b}+\mathcal{O}(x^{a-b-1}) \:\:\:\:\:\:\: 
\mbox{for $x\gg 1 $}
\eea 
(see Eqs.~(6.1.18) and~(6.1.47) of \cite{abramowitz72}).

\subsection{Recurrent Relations for spherical Bessel Functions} 
We use several recurrence relations for spherical Bessel functions  in
our derivations, most notably  
\be 
\frac{\ell+1}{x}j_{\ell}(x)+j'_{\ell}(x)=j_{\ell-1}(x)
\label{eq:AS-10.1.21} 
\ee 
and 
\begin{equation} 
\frac{\ell}{x}j_{\ell}(x)-j'_{\ell}(x)=j_{\ell+1}(x)~. 
\label{eq:AS-10.1.22} 
\end{equation}


\begin{thebibliography}{99} 
\bibitem{vachaspati01} T.  Vachaspati,  Phys. Rev. Lett. {\bf 87} 
  251302 (2001).  
\bibitem{brandenburg02} A.~Brandenburg, and E.~Blackman, 
astro-ph/0212019.
\bibitem{jedamzik98} K.~Jedamzik, V.~Katalini\'c, and A.V.~Olinto, 
Phys.\ Rev.\ D {\bf 57}, 3264 (1998). 
\bibitem{durrer00} R.~Durrer, P.G.~Ferreira, and T.~Kahniashvili, 
Phys.\ Rev.\ D {\bf 61}, 043001 (2000). 
\bibitem{grasso01} D.~Grasso and H.R.~Rubinstein, 
Phys.\ Rep.\ {\bf 348}, 163 (2001). 
\bibitem{mack02} A.~Mack, T.~Kahniashvili, and A.~Kosowsky, 
Phys.\ Rev.\ D  {\bf 65}, 123004 (2002). 
\bibitem{pogosian02}  L.~Pogosian, T.~Vachaspati, S.~Winitzki, 
Phys.\ Rev.\ D {\bf 65}, 3264 (2002). 
\bibitem{ahonen96} J.~Ahonen and K.~Enqvist, 
Phys.\ Lett.\ B {\bf 382}, 40 (1996). 
\bibitem{subramanian98b} K.~Subramanian and J.D.~Barrow, 
Phys.\ Rev.\ Lett.\ {\bf 81}, 3575 (1998). 
\bibitem{subramanian98a} K.~Subramanian and J.D.~Barrow, 
Phys.\ Rev.\ D {\bf 58}, 083502 (1998). 
\bibitem{hu96} W.~Hu and M.~White, 
Phys.\ Rev.\ D {\bf 56}, 596 (1997). 
\bibitem{caprini02} C.~Caprini and R.~Durrer, 
Phys.\ Rev.\ D {\bf 65}, 023517 (2002). 
\bibitem{Jcap}R.~Durrer and C.~Caprini, JCAP in print (2003)
  [{\tt astro-ph/0305059 }]. 
%\bibitem{seshadri01} T.R.~Seshadri and K.~Subramanian, 
%Phys.\ Rev.\ Lett.\ {\bf 87}, 101301 (2001). 
%\bibitem{subramanian02} K.~Subramanian and J.D.~Barrow, 
%MNRAS {\bf 57L}, 57S (2002). 
\bibitem{jackson75} J.D.~Jackson, {\it Classical Electrodynamics} 
(Wiley, New York, 1975). 
\bibitem{dk98} R. Durrer and M. Kunz, Phys. Rev. D {\bf 57}, R3199
  (1998). 
\bibitem{spergel} D.N. Spergel et al, astro-ph/0302209.
\bibitem{mark}M. Christensson, M. Hindmarsh and
  A. Brandenburg, Phys. Rev. E {\bf 64},  056405 (2001).
\bibitem{durrer98b} R.~Durrer and T.~Kahniashvili, 
Helv. Phys. Acta, {\bf 71}, 445 (1998). 
\bibitem{abramowitz72} M.~Abramowitz and I.~Stegun, 
{\it Handbook of Mathematical Functions}, Dover, New York (1972). 
%\bibitem{mukhanov92} V.F.~Mukhanov, H.A.~Feldman, and R.H.~Brandenberger, 
%Phys.\ Rep.\ {\bf 215}, 203 (1992). 
\bibitem{durrer94} R.~Durrer, Fundam.\ Cosm.\ Phys.\ {\bf 15}, 209 (1994). 
\bibitem{zaldarriaga97} M.~Zaldarriaga and U.~Seljak, 
Phys.\ Rev.\ D {\bf 55}, 1830 (1997). 
\bibitem{kamionkowski97b} M.~Kamionkowski, A.~Kosowsky, and 
A.~Stebbins,  Phys.\ Rev.\ D {\bf 55}, 7368 (1997). 
\bibitem{kosowsky96} A.~Kosowsky and A.~Loeb, 
Astrophys.\ J. {\bf 469}, 1 (1996). 
\bibitem{inflat}M.S.Turner and L.M. Widrow,  Phys. Rev. D {\bf 37},
        2743 (1988);\\
        B. Ratra,  Astrophys. J. Lett. {\bf 391} L1 (1992);\\
        W.D.Garretson, G.B. Field and S.M.Carroll, Phys. Rev. D 
        {\bf 46} 5346 (1992);\\
        O.Bertolami and D.F.Mota, Phys. Lett. B {\bf 455}, 96 (1999);\\
        A. Davis, K. Dimopoulos, T. Prokopec and O. T\"ornkvist,
        Phys. Lett.  {\bf B501}, 165 (2001).
\bibitem{stringcos} M. Gasperini, M. Giovannini and G. Veneziano, 
        Phys. Rev. Lett. {\bf 75}, 3796 (1995);\\
        D. Lemoine and M. Lemoine,  Phys. Rev. D {\bf 52}, 1955 (1995).
\bibitem{gradshteyn94} I.S.~Gradshteyn and I.M.~Ryzhik, 
{\it Table of Integrals, Series, and Products}, Academic Press, New
York and London (1965).
\end{thebibliography}
\end{document}